\documentclass{aa}  
\bibpunct{(}{)}{;}{a}{}{,} % to follow the A&A style
\usepackage{graphicx}
\usepackage{txfonts}
\usepackage{multirow}
\usepackage{soul}
\usepackage{needspace}
\graphicspath{ {.} }
\usepackage[usenames, dvipsnames]{xcolor}

\def\halpha{H$\mathrm{\alpha}$}
\newcommand{\CaII}{\ion{Ca}{ii}}

\newcommand{\MgII}{\ion{Mg}{ii}}
\newcommand{\MgIIk}{\ion{Mg}{ii}\,k}

\newcommand{\MgIIhk}{\ion{Mg}{ii}\,h\,\&\,k}

\def\kms{\hbox{km$\;$s$^{-1}$}} % A&A prefers km s-1 rather than km/s

\begin{document}

   \title{Ellerman bombs and UV bursts: \\ reconnection at different atmospheric layers}
   \titlerunning{Ellerman bombs and UV bursts}

   \author{Ada Ortiz
          \inst{1,2}
          \and
          Viggo H. Hansteen
          \inst{1,2}
          \and
          Daniel N\'obrega-Siverio
          \inst{1,2}
          \and
          Luc Rouppe van der Voort
          \inst{1,2}
          }

   \institute{Rosseland Centre for Solar Physics, University of Oslo, PO Box 1029 Blindern, 0315 Oslo, Norway \\
              \email{ada@astro.uio.no}
         \and
            Institute of Theoretical Astrophysics, University of Oslo, PO Box 1029 Blindern, 0315 Oslo, Norway \\
             }

   \date{Received / Accepted }

% \abstract{}{}{}{}{} 
% 5 {} token are mandatory

\abstract{The emergence of magnetic flux through the photosphere and into the outer solar atmosphere produces, amongst other dynamical phenomena, Ellerman bombs (EBs), which are observed in the wings of \halpha~and are due to magnetic reconnection in the photosphere below the chromospheric canopy. Signs of magnetic reconnection are also observed in other spectral lines, typical of the chromosphere or the transition region. An example are the UV bursts observed in the transition region lines of \ion{Si}{IV} and the upper chromospheric lines of \ion{Mg}{II}. In this work we analyze high cadence, high resolution coordinated observations between the Swedish 1-m Solar Telescope (SST) and the Interface Region Imaging Spectrograph (IRIS) spacecraft. \halpha~images from the SST provide us with the positions, timings and trajectories of EBs in an emerging flux region. Simultaneous, co-aligned IRIS slit-jaw images at 133 (C~{\sc ii}, transition region), 140 (\ion{Si}{IV}, transition region) and 279.6 (\ion{Mg}{II} k, core, upper chromosphere) nm, as well as spectroscopy in the far and near ultraviolet from the fast spectrograph raster, allow us to study the possible chromospheric/transition region counterparts of those EBs. Our main goal is to study the possible temporal and spatial relationship between several reconnection events at different layers in the atmosphere (namely EBs and UV bursts), the timing history between them, and the connection of these dynamical phenomena to the ejection of surges in the chromosphere. We also investigate the properties of an extended UV burst and their variations across the burst domain. Our results suggest a scenario where simultaneous and co-spatial EBs and UV bursts are part of the same reconnection system occurring sequentially along a vertical or nearly vertical current sheet. Heating and bidirectional jets trace the location where reconnection takes place. These results support and expand those obtained from recent numerical simulations of magnetic flux emergence.}


   \maketitle
%
%-------------------------------------------------------------------

\section{Introduction}
\label{intro}

The emergence of new magnetic flux from the solar interior into the photosphere and the regions above is a fundamental process as it facilitates the uplifting and renewal of magnetic flux in the solar atmosphere, and it is key player in the life cycle of the global magnetic field. A plethora of transient dynamical phenomena including surges, flares, jets, UV bursts, Ellerman Bombs (EBs) and coronal mass ejections (CMEs) -- to mention but a few -- occur when the newly emerged magnetic field interacts with the pre-existing (so-called ambient) field and/or with itself (see, e.g., \citealp{1995Natur.375...42Y},
\citealp{2011SSRv..159...19F},
\citealp{2011LRSP....8....6S},
\citealp{2013ApJ...771...20M}, 
\citealp{2016SSRv..201....1R}). 
The mentioned transient phenomena help the new field to rise and to push the old field aside by filling the chromosphere and corona, because they alleviate the mass that the rising field lines carry within. With every reconnection event, mass is ejected (or drains downward)
and that allows the magnetic field to attain coronal heights and form the long loops that form the active region corona \citep[see, e.g.,][]{2004ApJ...614.1099P}. It also constitutes a significant input of mass and energy, which are impulsively released to the upper solar atmosphere and even to the solar wind.

The interaction between flux systems depends on the {\it strength} of the field involved, on the {\it relative angle} between them, and on the {\it height} at which reconnection occurs \citep[see, e.g.,][]{1997ESASP...404..103S,2004A&A...426.1047A,2007ApJ...666..516G,2007ApJ...657L..53I,2015A&A...576A...4M}. For instance, EBs are produced when reconnection happens at low altitudes while X-ray jets occur when the interaction takes place at greater heights.

Some excellent reviews covering in depth the topic of flux emergence, both from an observational and from a numerical point of view, are those by \citet{2012RSPTA.370.3088A,2014LRSP...11....3C} and \citet{2014SSRv..186..227S}. In this work, we focus on two of the aforementioned transient dynamical phenomena related to flux emergence and reconnection, as well as on the possible relationship between them: EBs and UV bursts. 

EBs \citep{1917ApJ....46..298E} are defined by \citet{2015ApJ...812...11V} as "substantial brightenings of the extended wings of \halpha~without core brightening which, at sufficient angular and temporal resolution, show definite rapid-flame morphology when viewed from aside". These authors analyzed extensively the properties of EBs, their relationship to other phenomena, and their visibility in other wavelengths in a series of papers \citep{2011ApJ...736...71W,2013ApJ...774...32V,2015ApJ...812...11V}. In that series, they established that EBs are purely photospheric phenomena that trace the magnetic reconnection of strong opposite-polarity field concentrations in the low photosphere. \citet{2015ApJ...812...11V} made it clear that many of the EBs found in the literature are actually not real EBs, but "pseudo-EBs" that also appear bright in the wings of \halpha~due to radiation escaping from deeper layers rather than heating by photospheric reconnection. Quoting these authors, "care must be taken to ascertain that features that appear bright in an \halpha~wing are indeed EBs and not just facular brightenings —a warning already given by Ellerman (1917) himself". 

UV bursts \citep{2014Sci...346C.315P} are defined, according to the recent review by \citet{2018SSRv..214..120Y}, as compact ($\leq 2\arcsec$, typically $1\arcsec$), transient (lifetimes between tens of seconds to over an hour), intense (but not related to flares) events observed in image sequences sampling UV bands. In addition, their \ion{Si}{IV} line profiles present a complex shape with extended wings, large broadening and sometimes multiple peaks. UV bursts appear in locations where small magnetic features of opposite polarities collide, and do not generally show a co-spatial brightening in the coronal SDO/AIA channels. Different works consider different flavors of UV bursts, which may lead to different results. While \citet{2014Sci...346C.315P} analyze simple compact brightenings of about $1-2 \arcsec$ of diameter, the multi-wavelength study of \citet{2018ApJ...856..127G, 2019ApJ...871...82G} considers very extended brightenings of around $5 \arcsec$ in diameter that are longer-lived (three hours) than previously reported bursts and present coronal counterparts in all passbands of SDO/AIA. They conclude that the reconnection site occurs higher in the atmosphere than usually found in UV bursts, which could explain the coronal counterparts. The magnetic topology above UV bursts will determine its visibility at coronal levels. 

In recent times, there has been a debate in the literature about the relationship between these two particular phenomena -- if any -- and where in the atmosphere UV bursts form in relation to EBs. Many of the existing works that study this relationship advocate that UV bursts are hot pockets of gas originating very low in the solar atmosphere (even at upper photospheric heights) that would then be heated to transition region temperatures around $8\times 10^4$~K \citep[e.g.][]{2014Sci...346C.315P,2015ApJ...812...11V,2016A&A...593A..32G,2016ApJ...824...96T,2018ApJ...854..174T}. Using \ion{He} observations, and being the first to report on EB signatures in the \ion{He}{I} $D_3$ and \ion{He}{I} ~1083~nm lines, \citet{2017A&A...598A..33L} derive EB temperatures of order $T \sim 2\times 10^4 - 10^5$~K because they exhibit emission signals in neutral helium triplet lines. If the above studies are correct, current models of EBs need to be revised, as they are far from predicting such high photospheric temperatures. In fact models show that EBs cannot be reproduced if the photospheric temperature is too high 
\citep[e.g., ][]{2017ApJ...835L..37R, 2017ApJ...845..144H}. Recently \citet{2019A&A...627A.101V} found, from chromospheric spectral inversions, only a few thousand K temperature enhancements around the height of the temperature minimum in co-spatial EBs and UV bursts. However, no satisfactory results could be achieved for coupled inversions that included both the \ion{Si}{iv} and the chromospheric \ion{Ca}{ii} and \ion{Mg}{ii} lines at the same time. The inversion derived that the high temperatures in the lower atmosphere required to produce sufficient \ion{Si}{iv} emission resulted in unrealistic synthetic \ion{Ca}{ii} and \ion{Mg}{ii} profiles, suggesting that EBs do not achieve transition region temperatures.

\citet{2014Sci...346C.315P} could not conclude whether EBs and UV bursts were the same phenomena or not. \citet{2016ApJ...824...96T} claim that some of their UV bursts are also EBs and therefore form in the photosphere, which would be locally heated to $\sim 8 \times 10^4$ K. The argument is given based on \ion{O}{IV} emission, \ion{Mn}{I} absorption in the wings of \ion{Mg}{II} k, deep absorption \ion{Ni}{II} lines superimposed in the \ion{Si}{IV} 139.4 nm profiles, compact brightenings in the AIA~170~nm passband and enhanced emission in the \ion{Mg}{II} wings but not core. They note that some other UV bursts in their data sets may originate in the chromosphere and have no connection to EBs. 

Concerning the theoretical approach, \cite{2015ApJ...808..116J}, through 1D average atmospheric models, explained that the spectra of UV bursts have to have their origin, at least, $550$~km above the continuum photosphere. Recent 2.5D and 3D numerical simulations have found evidences that UV bursts are indeed formed in the chromosphere or above, and can possibly be the chromospheric counterparts of EBs.

In particular, by means of forward modeling of 2.5D magnetic flux emergence experiments, \cite{2017ApJ...850..153N} showed that UV burst-like \ion{Si}{IV} profiles are obtained in the current sheet between the emerged plasma and the preexisting ambient field at coronal heights of 5-6 Mm above the photosphere. This model was also used by \cite{2017ApJ...851L...6R} to synthesize \ion{Ca}{II}~K and provide theoretical support to the profiles obtained with CHROMIS observations at UV bursts.

With respect to 3D models, \citet{2017ApJ...839...22H} analyze a flux emergence scenario in which reconnection between colliding cold magnetized bubbles is triggered at different heights, such that EBs are purely photospheric and reach temperatures of $T \sim 8-9~\times 10^3$ K, UV bursts are a mid chromospheric phenomena (originating at 1.3 Mm in the case shown in the paper) with temperatures of $T \sim 7-8~\times 10^4$ K, and chromospheric microflares (at 1.8 Mm) reach $10^6$ K. However, in that model no co-located nor co-temporal EBs and UV bursts were reproduced. 

\citet[paper I henceforth]{2019A&A...626A..33H} takes this analysis further, using a 3D model similar to \citet{2017ApJ...839...22H}, but starting with a much stronger initial coronal field (2~Gauss at 10~Mm) oriented at a large angle to the direction of the injected emerging field. The initial model used for Paper I is a version of the publicly available ``enhanced network'' Bifrost model \citep{2016A&A...585A...4C} but where the numerical resolution has been increased by a factor of two, covering the same spatial extent, and having roughly the same magnetic topology. The stronger ambient field helps contain and confine the emerging field, such that large angle reconnection and powerful coronal heating is produced as the fields interact. The main result of Paper I is that EBs and UV bursts are produced, and occasionally these phenomena are found to be both {\it co-spatial and simultaneous}. This result is achieved through forward modeling, synthesizing, among 
other lines, the \halpha~and \ion{Si}{IV} 139.4 nm lines. This numerical experiment explains the relationship between EBs and UV bursts as due to a long vertical (or nearly vertical) current sheet that stretches from the photosphere to around 3000 km upwards where intensive reconnection and heating occurs along the entire current sheet. The EB is formed from the photosphere up to 1200~km above, while the UV burst is formed at heights between 700~km and 3~Mm above the photosphere. In other words, these two dynamical phenomena are part of the same system, but occurring at different heights along a reconnection {\it wall}.

With respect to the relation of EBs and UV bursts with other phenomena, both have been observed together with
surges. The EB-surge relation was reported already in the early 70s \citep{1973SoPh...32..139R}, and has been explored in more recent literature \citep[see, e.g.,][]{2011ApJ...736...71W,2013ApJ...774...32V,2013SoPh..288...39Y}; nonetheless, this relation does not seem to be common \citep{2013JPhCS.440a2007R}. A connection has been also established between UV bursts and surges: from the observational point of view \citep[e.g.,][]{2018ApJ...856..127G, 2019ApJ...871...82G} and from the theoretical perspective through numerical experiments \citep{2017ApJ...850..153N,2018ApJ...858....8N}. Both approaches show that those phenomena occur as a result of magnetic flux emergence from the solar interior. Taking into account the aforementioned findings, it is evident that an in-depth analysis is necessary to scrutinize the temporal and spatial relationship between EBs, UV bursts, surges and magnetic flux emergence.

Here we present multi-wavelength, high cadence, high resolution coordinated observations to study the possible relationship between reconnection events at different layers in the atmosphere, and in particular, the timing history between them. Our main goal is to study whether there is a temporal and spatial relationship between the appearance of an EB and the appearance of its associated UV burst. Of particular interest is the connection of these transient phenomena to the ejection of surges in the chromosphere. In addition we have investigated in detail the properties of one of the presented UV bursts, carefully examining the variations of these properties with position across the burst area. 

The layout of the paper is as follows: Section~\ref{obsdata} introduces our observations, the data reduction process and identification methods. Section~\ref{results} presents our results, which are then discussed and put into context in Section~\ref{discu}.

\section{Observations, data reduction and methodology}
\label{obsdata}

\subsection{Observations and data reduction}
\label{obs}

For the work presented here we have used coordinated observations between the Interface Region Imaging Spectrograph \citep[IRIS;][]{2014SoPh..289.2733D} and the Swedish 1-meter Solar Telescope \citep[SST;][]{2003SPIE.4853..341S} that were carried out during the last days of August and the beginning of September 2016 centered on AR 12585, which at that time was close to disk center. In particular, we are presenting observations from September 3rd, 4th, 5th and 6th as follows: September 3, center of the IRIS field-of-view (FOV) $(x,y)=(-561,44) \arcsec$, heliocentric angle $\mu$ = 0.80; September 4, center of the IRIS FOV $(x,y)=(-374,27) \arcsec$, heliocentric angle $\mu$ = 0.90; September 5, center of the IRIS FOV $(x,y)=(-161,24) \arcsec$, heliocentric angle $\mu$ = 0.98; and finally September 6, center of the IRIS FOV $(x,y)=(57,22) \arcsec$, heliocentric angle $\mu$ = 0.99.

The IRIS data sets correspond to the observing program with
OBS-ID 3625503135 carried out every day from 07:44:46 until 10:03:44 or to 10:38:06 UT depending on the day. These observations cover a FOV of $60 \arcsec \times 65 \arcsec$ for the slit-jaw images (SJI) and $5 \arcsec \times 60 \arcsec$ for the spectral raster. This observing program is a medium dense 16 step raster, which means that a $60 \arcsec$-long slit was shifted through 16 positions in the $x$ direction separated by $0\farcs33$, with a raster cadence of 20.8 s. The exposure time per slit position was 0.5 s, the IRIS pixel size is $0\farcs166$, and the FUV spectral bins were binned four times in order to increase the signal-to-noise level. The SJIs were obtained in three UV bands: at $133\,$nm (FWHM 5.5~nm, dominated by \ion{C}{II} lines and continuum), $140\,$nm (FWHM 5.5~nm, dominated by \ion{Si}{IV} lines and continuum), and $279.6\,$nm (FWHM 0.4~nm, centered on \ion{Mg}{II}~k) with a 10~s cadence.

With regards to the SST observations, we took spectroscopic images in the \ion{H}{I}~656.3~nm (\halpha) line, and spectropolarimetric images in the \ion{Ca}{II}~854.2~nm line and the \ion{Fe}{I}~$630.2$~nm line with the CRisp Imaging SPectropolarimeter \citep[CRISP;][]{2006A&A...447.1111S}. CRISP is a dual etalon Fabry-P\'erot interferometer mounted in telecentric configuration at the SST. Spectral profiles can be constructed for each pixel over the entire FOV by sequentially acquiring images over a given spectral window, typically at a rate of four spectral positions per second (or one spectral position in polarimetric mode). For this occasion \halpha~was sampled at 15 spectral positions, from $-0.15$~nm to 0.15~nm from the line center, with 8 exposures per wavelength position. The \ion{Ca}{II}~854.2~nm line was non-uniformly sampled at 21 positions in polarimetric mode, from $-0.175$ nm to 0.175 nm, taking 6 exposures per polarimetric state per wavelength position. Finally, the \ion{Fe}{I}~$630.2$~nm lines were sampled at 16 spectral positions in polarimetric mode with 6 exposures per polarimetric state per wavelength position. We usually performed short scans that contained only the \halpha~and \ion{Ca}{II}~854.2~nm lines, at a cadence of 20.2 s. Longer scans contained, in addition to the previous two, the \ion{Fe}{I}~$630.2$~nm lines at a cadence of 32.2 s. The SST pixel scale at 854.2~nm is $0\farcs057$ and the FOV was $58 \arcsec \times 58 \arcsec$.

The SST observations were reduced using the CRISPRED reduction pipeline \citep{2015A&A...573A..40D} which includes image reconstruction with the Multi-Object-Multi-Frame-Blind-Deconvolution technique \citep[MOMFBD,][]{2005SoPh..228..191V}. The seeing was moderate to good and the image quality benefited from the SST high order adaptive optics system \citep{2003SPIE.4853..370S}.

Aside from using the SST intensity narrowband filtergrams as such, we also constructed photospheric magnetograms from the \ion{Ca}{II}~854.2~nm observations. The formation height range of this line is quite large: while the core samples the mid-chromosphere, the far wings sample the photosphere, and we can add the Stokes V maps corresponding to the three outer spectral positions in both wings to obtain signed circular polarization maps that serve as a proxy for the photospheric magnetic field \citep[see, e.g.,][]{2014ApJ...781..126O}.

The temporal overlap between the IRIS and the SST data sets has the following durations: 2 hours and 14 minutes (September 3rd, from 07:49 to 10:03 UT), 1 hour and 9 minutes (September 4th, from 08:10 to 09:19 UT), 1 hour and 58 minutes (September 5th, from 07:45 to 09:44 UT) and 33 minutes (September 6th, from 07:47 to 08:20 UT). The alignment between the CRISP data and the IRIS observations was done in the following way: CRISP data was scaled down to the IRIS pixel scale ($0\farcs166$), followed by a cross-correlation of \ion{Ca}{II}~854.2~nm wing images with the SJI $279.6\,$nm channel. Table~\ref{table0} presents a quick overview of our observations. More details about the studied cases are given in Table~\ref{table1}.

\begin{table*}
\caption{Overview of the analyzed data sets. Columns are as follows: observation date, coordinates of the reconnection event in $\arcsec$, heliocentric angle, starting time of each reconnection event (UT, separated by comma if multiple brightenings), total duration of data set in minutes}             
\label{table0}      
\centering          
\begin{tabular}{| c | c | c | c | c | c |}     % 6 columns 
\hline
\rule{0pt}{3ex} 
Case & Observation date & Coordinates & $\mu$ angle & Start time & Duration of data set \\
 & & $(x,y) \arcsec$ & & (UT) & (min) \\
\hline
\rule{0pt}{2.3ex}  
1 & 3/9/2016 & $(x,y)_{EB}=(-555.7,47.7) \arcsec$ & $\mu$ = 0.80 & $t_{EB}$=08:13:30, 09:25:30 & 126 \\ 
 &  & $(x,y)_{UV}=(-556.5,47.9) \arcsec$ &  & $t_{UV}$=08:14:30, 09:27:30 & \\ 
\hline
\rule{0pt}{2.3ex}  
2 & 6/9/2016 & $(x,y)_{EB}=(58.3,8.1) \arcsec$ & $\mu$ = 0.99 & $t_{EB}$=07:59:17 & 21 \\  
  & & $(x,y)_{UV}=(57.7,8.6) \arcsec$ & & $t_{UV}$=08:05:00 & \\
\hline
\rule{0pt}{2.3ex}  
3 & 5/9/2016 & $(x,y)_{EB}=(-162.5,39.5) \arcsec$ & $\mu$ = 0.98 & $t_{EB}$=09:26:04 & 23 \\  
 & & $(x,y)_{UV}=(-162.5,39.8) \arcsec$ &  & $t_{UV}$=09:26:36 & \\  
\hline
\rule{0pt}{2.3ex}  
4 & 5/9/2016 & $(x,y)_{EB}=(-158.0,3.5) \arcsec$ & $\mu$ = 0.98 & $t_{EB}$=08:16:36 & 17 \\  
\hline
\rule{0pt}{2.3ex}  
5 & 3/9/2016 & $(x,y)_{EB}=(-563.4,25.5) \arcsec$ & $\mu$ = 0.80 & $t_{EB}$=08:50:48, 08:58:13 & 68 \\  
 &  & $(x,y)_{UV}=(-562.9,25.5) \arcsec$ &  & $t_{UV}$=08:51:08, 08:59:34  & \\ 
\hline
\rule{0pt}{2.3ex}  
6 & 4/9/2016 & $(x,y)_{UV}=(-369.3,41.0) \arcsec$ & $\mu$ = 0.90 &  $t_{UV}$=08:10:46, 08:58:59 & 69 \\  
\hline
\rule{0pt}{2.3ex}  
7 & 5/9/2016 & $(x,y)_{UV}=(-160.7,37.1) \arcsec$ & $\mu$ = 0.98 & $t_{UV}$=08:53:00 & 29 \\  
\hline
\rule{0pt}{2.3ex}  
8 & 5/9/2016 & $(x,y)_{UV}=(-160.3,10.8) \arcsec$ & $\mu$ = 0.98 & $t_{UV}$=09:08:53, 09:15:20 & 29 \\  
\hline                  
\end{tabular}
\end{table*}

To investigate any possible lower chromospheric or coronal response to the reconnection events presented here, we also made use of SDO/AIA imagery \citep{2012SoPh..275...17L} from the 170~nm (temperature minimum, $\sim 5\times 10^3$~K), 30.4 nm (chromosphere, transition region, $5\times 10^4$~K), 17.1 nm (upper transition region, $\sim 6.3\times 10^5$~K), and 19.3 nm (corona, $3\times 10^6$~K) channels, as well as of photospheric SDO/HMI magnetograms \citep{2012SoPh..275..207S}.

\subsection{EB identification in \halpha~spectra and filtergrams}
\label{EB_iden}

For the identification of our EBs, we used the SST blue wing \halpha\ filtergrams at $-0.1$~nm from line center. We have made extensive use of the widget-based tool CRISPEX \citep{2012ApJ...750...22V} in order to identify the EBs. Once a bright patch was visually identified in the \halpha~wing filtergram we followed a similar approach to \citet{2013ApJ...774...32V} to confirm whether the specific brightening was an actual EB or not. We then looked at the \halpha~profiles of those bright pixels and used an intensity threshold when inspecting their spectra, so that pixels whose wings (both the blue and the red wing) were 150\% or more of the average intensity over the whole FOV were selected. In addition, the core of the line should not present any increase in intensity. After these requirements were imposed, we confirmed (when inspecting the photospheric magnetograms in the \ion{Ca}{II}~854.2~nm line) that the selected EB locations were, indeed, sites of merging between magnetic fields of opposite polarities. This procedure gave us the initial location of the EB, which was chosen as the pixel with highest intensity within the bright patch complying with the mentioned requirements.

A $15 \arcsec \times 15 \arcsec$ box was then selected around the initial position of the EB in the 2D \halpha~wing images. For the subsequent timesteps, the location of the EB was automatically inferred by calculating the centroid of those pixels inside the box above a certain intensity threshold. This threshold was set to 6500 counts in the \halpha~wing filtergrams. If no pixel was above the threshold at a particular timestep, the position of the EB was taken as the position in the previous timestep. The intensity given for the EB at a given time was calculated as the average of the fifty brightest pixels inside the box. This procedure thus avoids considering pixels inside the box that did not have an intensity above the minimum set and thus were not part of the EB. The \halpha~and \ion{Ca}{II} light-curves presented in Figures~\ref{fig_caso2_evtemp}, \ref{fig_caso3_evtemp}, \ref{fig_caso4_evtemp} and \ref{fig_caso7_evtemp} follow this method.

Movies of \halpha~$-0.1$~nm showing the temporal evolution and trajectories of the selected EBs can be found as supplemental materials to this article.

\subsection{UV burst identification in IRIS SJIs and spectra}
\label{UV_iden}

For the identification of UV bursts, we used the \ion{Si}{IV} $140\,$nm SJIs and spectra, following the definition of \citet{2014Sci...346C.315P}. In those SJIs, we identified small ($\sim 1 \arcsec$), compact, roundish transient brightenings with substantial emission in the \ion{Si}{IV} spectral lines observed by IRIS, whose spectra show very wide and complex non-Gaussian profiles on which deep absorption blends of lower metal ionization stages are superimposed. This definition gave us the initial location of the UV burst. Similar to the EB identification, we chose the pixel within the UV patch that presented a higher \ion{Si}{IV} intensity. 

Again a $15 \arcsec \times 15 \arcsec$ box was then selected around the initial position of the burst in the \ion{Si}{IV} $140\,$nm SJIs. Automatic detection of the UV burst in subsequent timesteps was carried out in the same way as for EBs, i.e., calculating the centroid of pixels inside the box above a certain intensity threshold, which in this case was set to 12 counts for the $140\,$nm SJIs. The intensity of the UV burst at a given timestep was taken as the average of the fifty brightest pixels inside the box. Whenever there was a South Atlantic Anomaly (SAA) event, the location of the UV burst was frozen at the position of the last timestep not affected by the SAA in order to avoid any spurious jumping of our detection box. The FUV and NUV light-curves in Figures~\ref{fig_caso2_evtemp}, \ref{fig_caso3_evtemp}, \ref{fig_caso4_evtemp} and \ref{fig_caso7_evtemp} follow this method.

Movies of \ion{Si}{IV} $140\,$nm and \ion{Mg}{II}~k $279.6\,$nm SJIs showing the temporal evolution and trajectories of the identified UV bursts can be found as supplemental material to this article.

\subsection{Time-sliced spectra}
\label{how_timesliced}

As part of the spectral analysis presented here we plot the spectra of EBs and UV bursts as a function of time (time-sliced spectra). The purpose is to show the temporal evolution of their spectra. The spectra shown for each timestep is that of the centroid calculated in Sects.~\ref{EB_iden} and \ref{UV_iden} at that given timestep, i.e., we follow the trajectory of the EB or UV burst as previously explained and show its spectra at every given timestep. Since coexisting EBs and UV bursts follow different trajectories (albeit closely related), the spectra shown in the time-sliced spectra plots for EBs is not derived exactly at the same pixel than for its associated UV burst. In the case where only an EB was present with no associated UV brightening, both the SST and the IRIS spectra have been plotted at the same pixel, i.e., that given by the trajectory of the EB. The same was done for the case of a UV burst only.

\section{Results}
\label{results}

We have analyzed a total of eight cases of transient brightenings associated with magnetic reconnection in the flux emergence region present within AR 12585, for the period September 3-6, 2016. The criteria for selecting those eight cases was that most of them should be located within the slit positions of the IRIS spectrograph, so that UV spectra -- and not just visible or infrared spectra from the SST -- were also available. Six out of the eight cases lie within the FOV of the IRIS slit spectrograph. The remaining two cases do not fall within the IRIS slits, but we chose them for their particular interest (both are long lived brightenings).

Four of the eight studied cases are presented in more detail in the subsections below as they are representative of the whole sample, and they are moreover covered by the IRIS raster. 

Table~\ref{table1} summarizes the characteristics of the eight cases of transient brightenings studied here. It is worth noting that some of the cases presented here have been examined with other purposes in the literature. Case 1 has been analyzed in \citet{2017ApJ...850..153N,2017ApJ...851L...6R} and \citet{2019A&A...627A.101V}, while \citet{2017ApJ...851L...6R} and \citet{2019A&A...627A.101V} focus part of their study in timesteps neighboring our case 3.

\begin{table*}[h!]
\caption{Characteristics of the eight transient brightenings presented in this work. Columns are as follows: event happens at the slit positions of the IRIS spectrograph, type of brightening, duration of each reconnection events (in minutes, first number refers to the EB, second number to the \ion{Si}{IV} UV burst, separated by a comma), delay of the UV burst respect to the EB (in minutes, only for cases where both co-exist), presence of an associated surge, existence of \ion{Mg}{II} triplet in emission, existence of \ion{Mn}{I} absorption lines superimposed in the blue wing of \ion{Mg}{II} k profiles, existence of \ion{Ni}{II} absorption lines superimposed in the \ion{Si}{IV} 139.4 nm profiles, coronal response in the AIA channels and brightening in the AIA 170~nm channel.}             
\label{table1}      
\centering          
\begin{tabular}{| c | c | c | c | c | c |}     % 6 columns 
\hline
\rule{0pt}{3ex} 
Case &  IRIS slit & Type of & Duration & Delay of UV & Surge \\
 &  & brightening & (min) & burst (min) & \\
\hline
\rule{0pt}{2ex}  
1 &  no & EB + UV burst & 31 + 38\tablefootmark{a}, 25.5 + 36\tablefootmark{b} & 1 & yes \\ 
2 &  yes & EB + UV burst & 21, 15 & 6 & yes \\  
3 &  yes & EB + UV burst & 22, 7.5 & 0.5 & yes \\  
4 &  yes & EB & 9 & -- & no \\  
5 &  yes & EB + UV burst & 6, 4.5\tablefootmark{c} & 0.5 & yes \\  
6 &  no & UV burst & 47+19\tablefootmark{b} & -- & yes \\  
7 &  yes & UV burst & 6+7+1\tablefootmark{b} & -- & yes \\  
8 &  yes & UV burst & 7+6+2\tablefootmark{b} & -- & yes \\  
\hline                  
\end{tabular}
\end{table*}

\begin{table*}[h!]
\centering
\begin{tabular}{| c | c | c | c | c | c |}     % 6 columns 
\hline
\rule{0pt}{3ex} 
Case & \ion{Mg}{II} triplet & \ion{Mn}{I} absorption on & \ion{Ni}{II} absorption & Coronal response & AIA 170~nm \\
 & emission & \ion{Mg}{II} k blue wing & on \ion{Si}{IV} 139.4 nm & in AIA & brightening  \\
\hline
\rule{0pt}{2ex}  
1 & -- & -- & -- & no & yes \\ 
2 & yes & yes\tablefootmark{d} & yes & no & yes \\  
3 & yes & yes\tablefootmark{d} (weak) & yes & no & yes \\  
4 & no & no & no & -- & -- \\  
5 & yes (weak) & no & no & no & yes \\  
6 & -- & -- & -- & yes & yes \\  
7 & no & yes (very weak) & no & no & yes \\  
8 & no & no & yes & no (yes in 30.4 nm) & no \\  
\hline                  
\end{tabular} \\
\tablefoot{ \\
\tablefoottext{a}{This EB presents several brightenings. These numbers refer to each brightening.}\\
\tablefoottext{b}{This UV burst presents several brightenings. These numbers refer to each brightening.}\\
\tablefoottext{c}{These events present several intensity peaks. These values refer to the peaks with higher intensity.} \\
\tablefoottext{d}{During some timesteps.}
}
\end{table*}

\subsection{Coexistence of an EB and a UV burst: September 6, 2016}
\label{caso2}

\begin{figure*}[h!]
\centering
\includegraphics[width=0.83\textwidth]{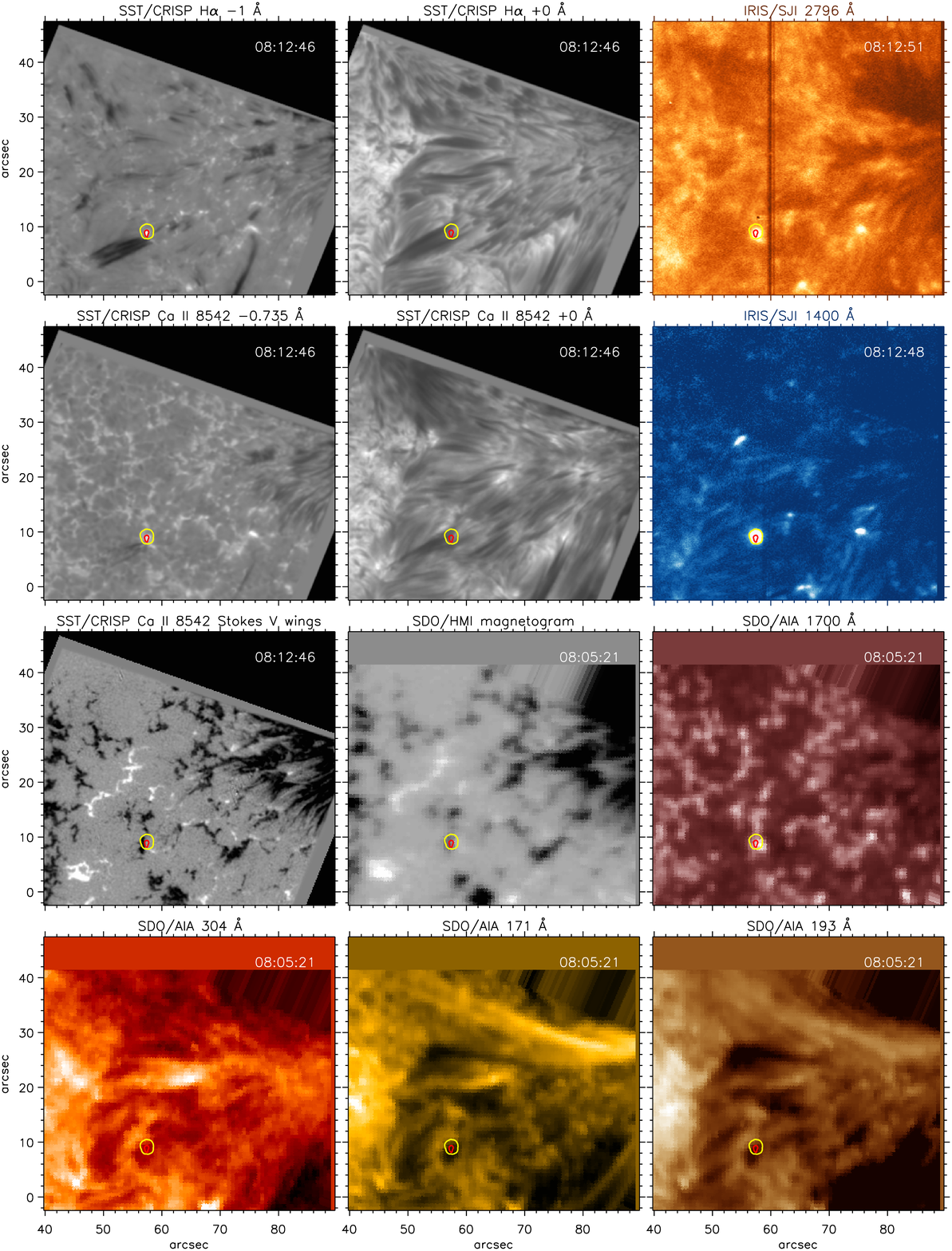} 
\caption{Mosaic of images from SST, IRIS and SDO for September 6, 2016. At this particular time both an EB and a UV burst co-existed temporally and spatially. Upper row (from left to right): SST \halpha~wing at $-0.1$~nm, SST \halpha~core, and IRIS SJI $279.6\,$nm. Second row: SST \ion{Ca}{II}~854.2~nm wing at -0.0735 nm, SST \ion{Ca}{II}~854.2~nm core and IRIS SJI $140\,$nm. Third row: SST Stokes V map in the wings of \ion{Ca}{II}~854.2~nm, SDO/HMI magnetogram and SDO/AIA 170~nm. Fourth row: SDO/AIA 30.4~nm, SDO/AIA 17.1~nm and SDO/AIA 19.3~nm. Red and yellow contours in all the panels highlight the location of the EB and UV burst, respectively. The Stokes V map in the wings of \ion{Ca}{II}~854.2~nm and the HMI photospheric magnetogram show that both reconnection events happen at a site of merging of opposite magnetic polarities. The AIA/SDO images do not show any coronal counterpart of these reconnection events, but a diminished intensity due to absorption by the surge.}
\label{fig_caso2_mos}
\end{figure*}

On 6 September 2016, between 07:59 and 08:20 UT, an EB and a UV burst coexisted spatially and temporally. The initial position of the EB was $(x,y)=(58.3,8.1) \arcsec$ at 07:59:17 UT, while the initial position of the UV burst was $(x,y)=(57.7,8.6) \arcsec$ starting at 08:05:00 UT. This pair of brightenings are presented in Figure~\ref{fig_caso2_mos}, which shows a mosaic of images from the SST, IRIS and SDO observatories at 08:12 UT (except for the SDO imagery, which is taken at 08:05:21 UT). This FOV is the site of vigorous flux emergence happening South-East of the sunspot partially seen in the upper-right part of the panels. 

The location where the brightenings occur is the site of merging of magnetic fields of opposite polarities, as shown by the Stokes V maps in the wings of \ion{Ca}{II}~854.2~nm and the HMI magnetograms. This is common for the eight cases presented here. The positive and negative polarities seem to interact intermittently. 

In the panels of Figure~\ref{fig_caso2_mos} the EB has been highlighted with a red contour (see Sect.~\ref{EB_iden} for details on the identification process), while the UV burst has been highlighted with a yellow contour (Sect.~\ref{UV_iden} gives details on the process followed to identify UV bursts). The UV burst is evident in both the IRIS SJI $140\,$nm and $279.6\,$nm. In these images the EB reaches a maximum diameter of around $1.5 \arcsec$ with a roundish or elongated shape during its lifetime. The UV burst presents a peak diameter of around $4.5 \arcsec$ with a predominantly circular shape. During the time of their co-existence the UV burst lies exactly on top of the EB, with the latter occupying always the center of the UV burst extension. 

The \halpha\ wing images show a surge shooting out diagonally from the lower-left side of the EB at around 08:06 UT that lasts for 13 minutes, almost until the end of the observations. The surge is a very dynamical structure with a maximum length of 16$\arcsec$ (reached at 08:14:27 UT) in the \halpha\ $-0.1$~nm images. The surge can also be observed in the \ion{Ca}{II}~854.2~nm panels, both in the wings and the core of the line. The temporal evolution of the surge can be seen in the \halpha~and \ion{Ca}{II}~854.2~nm movies presented as supplemental material. 

While the AIA 170~nm panel of Fig.~\ref{fig_caso2_mos} shows a brightening at the common position of the EB and UV burst, the transition region and coronal wavelengths do not show any signature associated to those reconnection events. What these high temperature EUV diagnostics show instead is indeed a darkening roughly coinciding with the surge. The cold dense material of the surge is opaque due to the presence of neutral hydrogen and helium and absorbs eventual radiation from the reconnection events not allowing any generated radiation in these EUV bands to escape.

\begin{figure}[h!]
\centering
\includegraphics[width=0.5\textwidth]{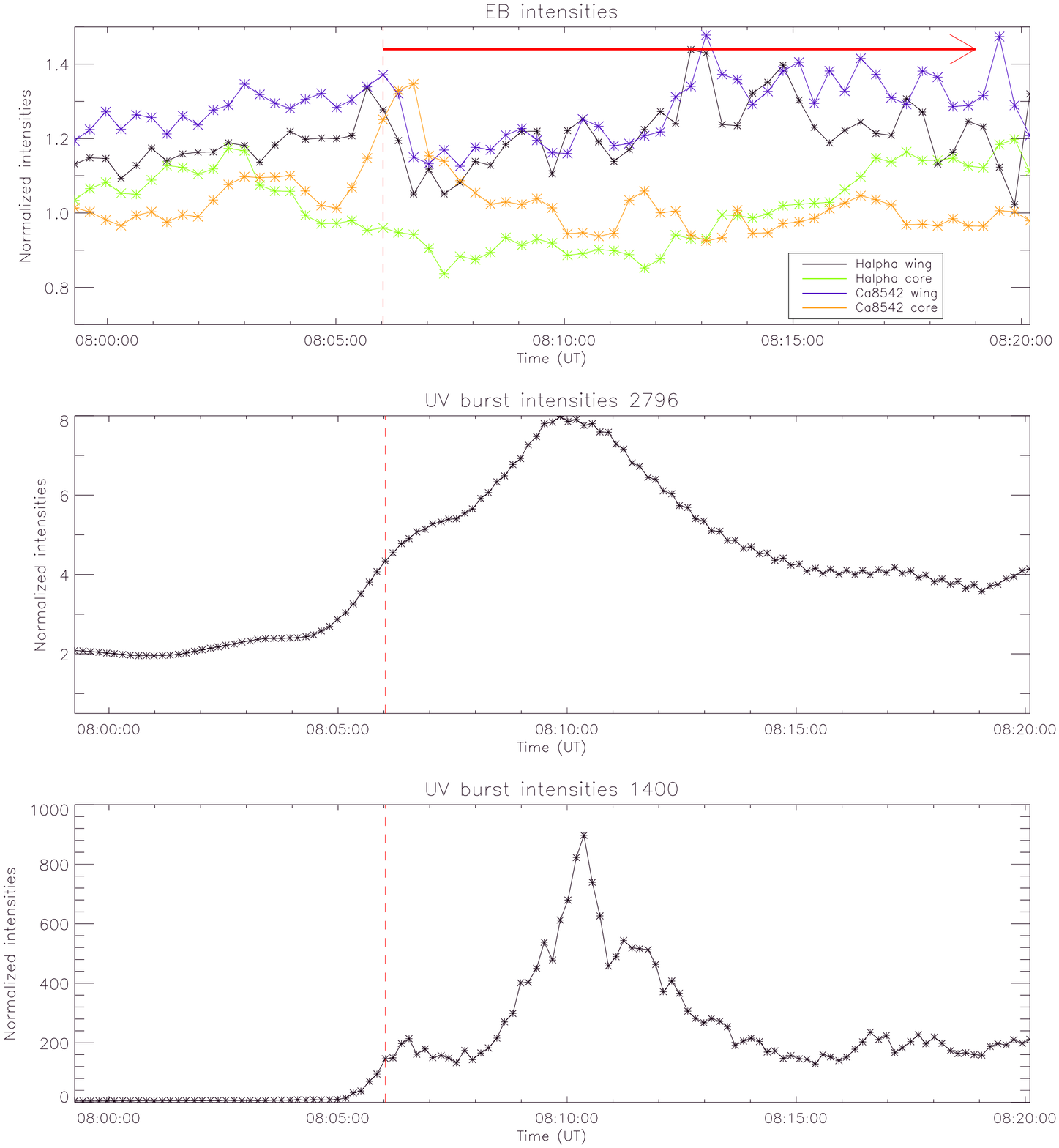}  
\caption{Intensity as a function of time during the duration of the co-spatial and co-temporal EB and UV burst brightenings for September 6, 2016. Upper panel: average SST \halpha\ (wing and core, black and green respectively) and \ion{Ca}{II}~854.2~nm (wing and core, blue and orange respectively) intensities in a box enclosing the EB; middle panel: average IRIS SJI $279.6\,$nm intensity in a box enclosing the UV burst; lower panel: average IRIS SJI $140\,$nm intensity in a box enclosing the UV burst. The vertical dashed line and the red arrow mark the onset and time-life of the surge.}
\label{fig_caso2_evtemp}
\end{figure}

Figure~\ref{fig_caso2_evtemp} presents the temporal evolution of the EB and UV burst intensities at several wavelengths for the duration of their coexistence. The light-curves have been calculated as detailed in Sects.~\ref{EB_iden} and \ref{UV_iden}. The upper panel shows SST intensities for \halpha\ $-0.1$~nm, \halpha\ line center, \ion{Ca}{II}~854.2~$-0.0735$~nm and line center. Both the \halpha\ intensities in the wing and the core of the line show a decrease at the time when the surge is totally covering the EB or is more active. The middle panel shows the intensity derived from the IRIS SJIs $279.6\,$nm images, while the lower panel presents the intensity obtained from the SJI $140\,$nm images. 

The EB starts at 07:59 UT and it is still present at the end of our observations, lasting at least for 21 minutes (see Table~\ref{table1}). The UV brightening however appears 6 minutes later, at 08:04:30 UT in the SJI $279.6\,$nm and at 08:05:00 UT in the SJI $140\,$nm images. At the end of the data set it is still well visible in both wavelengths, thus existing for at least 15 minutes. Its intensity peaks at 08:10:00 UT in \ion{Mg}{II} (middle panel) and 30 s later in \ion{Si}{IV} (lower panel). The EB intensity does not show substantial variations (only a maximum at 08:05:41 and 08:12:46 UT), while the UV burst intensity increases impulsively only taking 5 minutes from its appearance in the FOV until it reaches its maximum intensity. 

Figure~\ref{fig_caso2_evtemp} reveals a bursty behavior specially in the \ion{Si}{IV} intensity. This is common to the other UV bursts analyzed here. The timescale of this intermittent behavior in our observations is of the order of a few minutes (well sampled by our IRIS observations at a 10 s cadence), varying between three to ten minutes depending on the case. Similar variations have been reported by \citet{2018SSRv..214..120Y} and \citet{2018ApJ...856..127G}. The light curves of the \ion{Mg}{II} SJIs have a slightly smoother behavior when compared with the temporary evolution of the \ion{Si}{IV} SJIs as was also observed by \citet{2018ApJ...856..127G}.

\begin{figure*}
\centering
\resizebox{0.49\hsize}{!}{\includegraphics{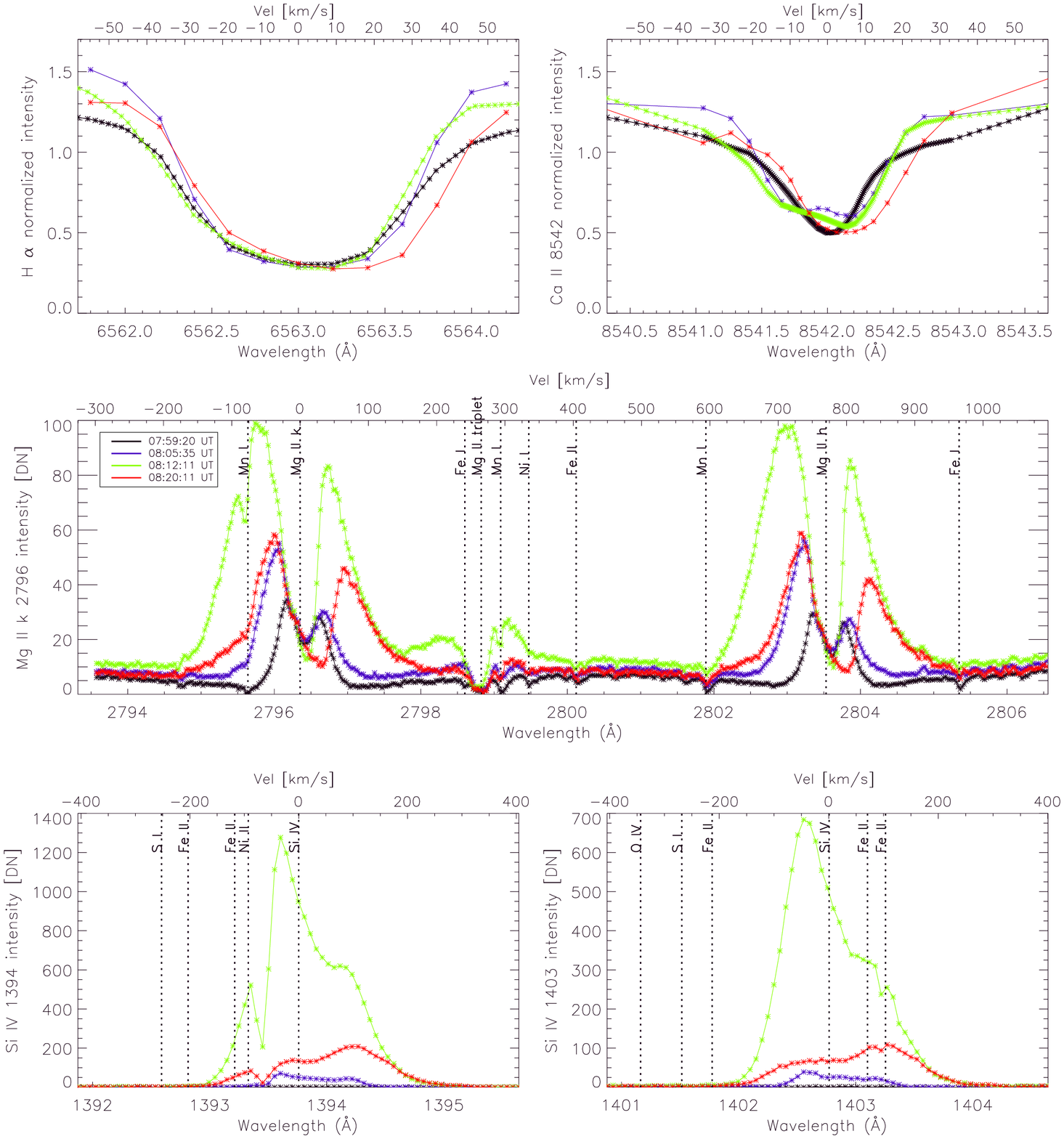}}
\resizebox{0.49\hsize}{!}{\includegraphics{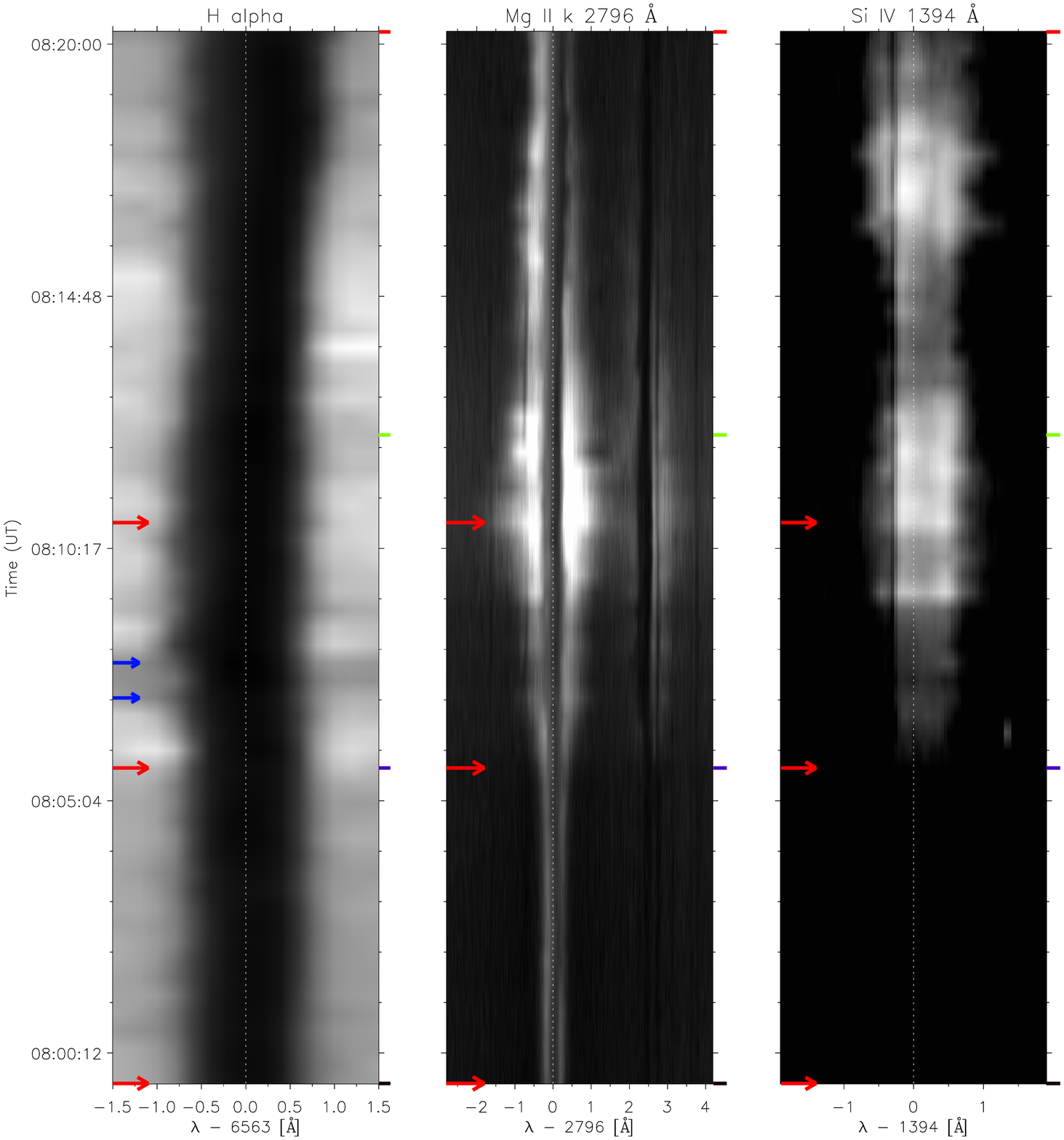}}
\caption{Left panel: Spectra of different lines for September 6, 2016 at specific times (see legend) during the evolution of the EB and the UV burst. Upper row: \ion{H}{I}~656.3~nm and \ion{Ca}{II}~854.2~nm respectively; mid row: \ion{Mg}{II}~k and h lines at $279.6\,$~nm; lower row: \ion{Si}{IV} lines at 139.4 and 140.3~nm respectively. The wavelengths at rest of some reference lines are indicated by vertical dotted lines. Right panel: time-sliced spectra, also for September 6 2016, for \ion{H}{I}~656.3~nm (left), \ion{Mg}{II}~k and triplet lines (center) and the \ion{Si}{IV} 139.4~nm line (right). The \ion{Ni}{II} absorption line can be seen superimposed on the left wing of \ion{Si}{IV} 139.4~nm during the whole duration of the UV brightening. Red and blue arrows mark specific moments in the evolution of the brightenings (see text for details). The colored marks pinpoint to the same times that have been highlighted in the left panel. We have applied a gamma adjustment to the \ion{Si}{IV} color table.}
\label{fig_caso2_spectra+timespectra}
\end{figure*}

Figure~\ref{fig_caso2_spectra+timespectra} presents spectral profiles for different SST and IRIS lines as well as their temporal evolution. In particular, the left panel shows the profiles for \halpha~and \ion{Ca}{II}~854.2~nm (top), \ion{Mg}{II}~k and h lines (middle), and the \ion{Si}{IV} lines at 139.4 and 140.3~nm respectively (bottom), for four selected time steps. These time steps are representative of the beginning of the EB (07:59 UT, black line, start of the observations), the beginning of the UV burst (08:05 UT, blue line), the peak of the UV burst diameter (08:12 UT, green line) and finally the end of the observations (08:20 UT, red line). The panel on the right is a time-sliced spectra showing the temporal evolution of three spectral lines: SST \halpha, IRIS \ion{Mg}{II}~k and triplet lines and IRIS \ion{Si}{IV} 139.4~nm. We have followed the procedure detailed in Sect.~\ref{how_timesliced}. Red and blue arrows in the time-sliced spectra pinpoint to specific moments in the evolution of the brightenings. The lower red arrow marks the beginning of the EB and the observations themselves. The middle red arrow points at the beginning of the UV burst. The top red arrow highlights the moment of maximum diameter of the UV burst. The blue arrows (08:06:45 to 08:07:30 UT) enclose the time of maximum \halpha~intensity drop due to the passage of the surge hiding the EB from the observer's view (this can also be clearly seen in the \halpha~movies for September 6, 2016 offered in the supplementary material). 

As shown by the top left panel of Figure~\ref{fig_caso2_spectra+timespectra}, the EB is brighter (i.e., has a higher increase in the \halpha~wings) at around 08:05 UT  -- right before the surge covers the EB -- with a similar behavior in the wings of \ion{Ca}{II}~854.2~nm. After the intensity decrease at 08:06:45 UT in the wings of \halpha, a recovery of the EB intensity follows until 08:13 UT. The UV burst, on the other hand, reaches a peak in diameter and brightness between 08:09 and 08:13 UT. This is evident in the \ion{Mg}{II}~k and h profiles at $279.6\,$~nm and the \ion{Si}{IV} profiles at 139.4 and 140.3~nm at 08:12 UT in the left panels of Figure~\ref{fig_caso2_spectra+timespectra} (green line). In the case of the \ion{Si}{IV} profiles the intensity increases by one order of magnitude when compared to any other time steps. In the case of the \ion{Mg}{II} profiles, the amplitude of the k2v and k2r peaks doubles at 08:12 UT relative to the other times.

An inspection of the \ion{Mg}{II}~k and h profiles, as well as the time-sliced spectra, makes evident the presence of the \ion{Mg}{II} triplet in emission for about 9 minutes around the time of maximum diameter and brightness of the UV burst. According to \citet{2015ApJ...806...14P} the triplet comes into emission after a temperature increase in the lower chromosphere of more than 1500 K. The \ion{Mg}{II} triplet in emission is very evident in the center-left panel of Figure~\ref{fig_caso2_spectra+timespectra} at 08:12 UT. The presence of the \ion{Ni}{II} absorption line is also evident when inspecting the \ion{Si}{IV}~139.4~nm profiles (left panel) and time-sliced spectra (right panel). The deep superimposed on its left wing is well visible in the \ion{Si}{IV} profile at 08:12 UT (green line) due to the increased intensity of that profile. However, the time-sliced spectra reveals that the absorption is present for the whole duration of the burst at that particular point and not only at the time of maximum brightness.

\subsubsection{A closer view of the UV burst: properties}
\label{UVburst}

In this subsection we analyze in more detail the properties of the UV burst and its variations with position within the burst area. 

This UV burst has an approximate extension of $2 \arcsec \times 2 \arcsec$ --reaching almost $2 \arcsec \times 3 \arcsec$ in its period of maximum extension-- when seen in the spectral rasters of IRIS. It has a roundish shape when observed in the $140\,$nm IRIS SJI (which maintains throughout its lifetime) but appears slightly smaller when seen in the spectrograph rasters. The \ion{Si}{IV} 140.3~nm profiles observed within the UV burst area come in a variety of shapes: clearly double-peaked, clearly singled-peaked or a mixture of both, i.e., mainly single-peaked with a lesser second component in one of the wings.

\begin{figure}
\centering
\resizebox{1.1\hsize}{!}{\includegraphics{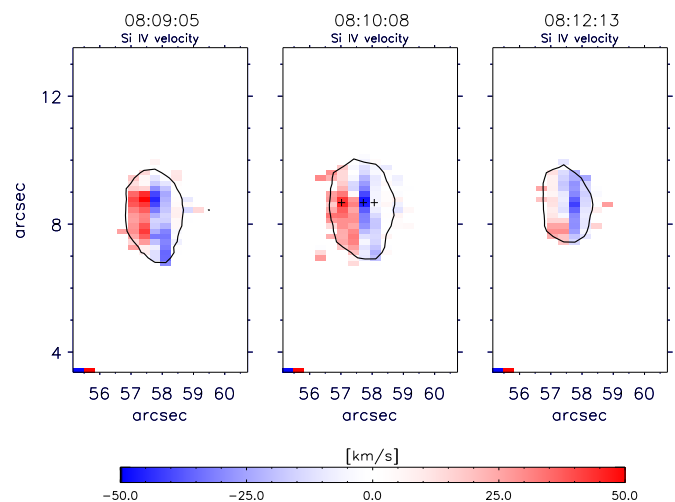}}
\caption{Maps of the \ion{Si}{IV} 140.3~nm velocity for three timesteps: from left to right, 08:09:05, 08:10:08 and 08:12:13 UT. The timesteps have been chosen around (before, during and after) the maximum intensity peak of the burst. The time of maximum intensity can be found in Figure~\ref{fig_caso2_evtemp} slightly after 08:10 UT. A single Gaussian has been fitted to the \ion{Si}{IV} 140.3~nm profiles (but see text for additional explanations). The crosses in the central panel mark the positions of the pixels whose profiles are shown in Figure~\ref{fig_caso2_si4_profiles}. The black contours delimit the boundaries of the UV burst for the different timesteps by imposing a threshold to the \ion{Si}{IV} intensity.}
\label{fig_caso2_si4_vel}
\end{figure}

In Figure~\ref{fig_caso2_si4_vel} we present the velocity maps derived from \ion{Si}{IV} 140.3~nm profiles corresponding to three timesteps around the maximum peak of intensity of the UV burst. The moment of maximum intensity of the UV burst can be seen in the middle and lower panels of Figure~\ref{fig_caso2_evtemp}. In particular we have chosen the following times: before the maximum at 08:09:05 UT, during the maximum at 08:10:08 UT and after the maximum at 08:12:13 UT. The black contours delimit the boundaries of the UV burst by imposing a threshold of 20 counts to the \ion{Si}{IV} raster intensities. The velocities represented in Figure~\ref{fig_caso2_si4_vel} have been derived by fitting a single Gaussian to the \ion{Si}{IV} profiles. While several profiles within the burst area are double-peaked as mentioned before -- and therefore could be very well fit using a double Gaussian -- the same double Gaussian fit works extremely poorly for those pixels who are single-peaked or a hybrid between single and double peaks. Therefore we opted for using a single Gaussian fit through all the UV burst domain in order to construct the maps shown in Figure~\ref{fig_caso2_si4_vel} at the expense of obtaining more conservative velocities in those pixels which show double-peaked profiles (thus not properly represented by a single Gaussian fit). The maximum velocities yielded by this single fit method are $\pm 50$ \kms (with the exception of a few pixels yielding -75 \kms). The pixels outside of the UV brightening have a very small signal-to-noise ratio and therefore no function can be properly fitted. Only pixels with peak intensity above $2.5\sigma$ have been included in the fits. 

Realizing that the velocities yielded by the fit of a single Gaussian are quite conservative in many of the burst pixels (especially in the double-peaked ones), we have individually fitted a double Gaussian to some of those pixels which are more accurately represented by this function. In this case the velocities reach up to -50 \kms for the blueshifted component and up to 100 \kms for the redshifted component, both at 08:09:05 and 08:10:08 UT. Such high values are not reached at 08:12:13 UT as this timestep shows lesser velocities.

Note that the burst domain is clearly divided into three different zones (redshifted velocities, blueshifted velocities and zero velocity) at all times, as the observed \ion{Si}{IV} profiles in those zones are redshifted, blueshifted or at rest respectively. The redshifted region is centered around $57 \arcsec$ in the $x$-direction with a width of roughly $1 \arcsec$, while stretching some $2\arcsec$ in the $y$-direction from $7.2 \arcsec$ to $9.2 \arcsec$. The blueshifted region lies $1 \arcsec$ to the right with a width of slightly less than $1\arcsec$. Initially the extent of the blueshifted region is slightly longer than the redshifted region, stretching from $6.8\arcsec$ to $9.2\arcsec$, but at later times the red- and blueshifted regions have a very similar extent. The \ion{Si}{IV} velocities are highest before and during the peak of maximum intensity, but decrease significantly at the time of the last panel at 08:12:13 UT.

\begin{figure}
\centering
\resizebox{\hsize}{!}{\includegraphics{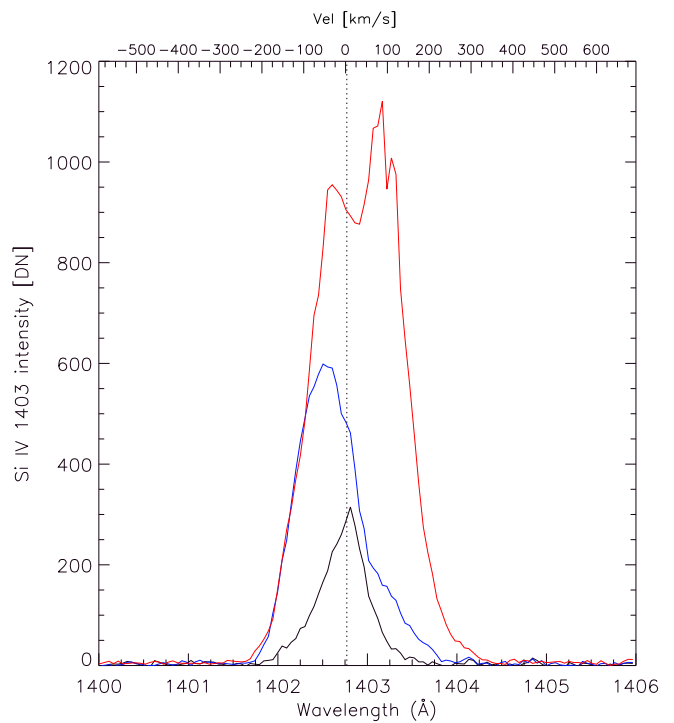}}
\caption{\ion{Si}{IV} 140.3~nm profiles corresponding to the locations marked by a cross in Figure~\ref{fig_caso2_si4_vel} at 08:10:08 UT (central panel). The red curve corresponds to the pixel marked by a cross within the red patch of the burst, the blue curve corresponds to the pixel marked by a cross within the blue patch of the burst, while the black curve corresponds to the pixel highlighted with a cross within the white patch of the burst in the central panel of Figure~\ref{fig_caso2_si4_vel}.}
\label{fig_caso2_si4_profiles}
\end{figure}

Representative \ion{Si}{IV} profiles for each of the above mentioned zones are given in Figure~\ref{fig_caso2_si4_profiles}. This figure shows the \ion{Si}{IV} profiles corresponding to those pixels marked by a cross in the central panel of Figure~\ref{fig_caso2_si4_vel} and showcases the variety of shapes found across the burst domain. The red curve is a double-peaked redshifted profile and corresponds to the pixel marked by a cross in the redshifted zone of the UV brightening. The blue curve is a blueshifted single-peaked profile with a smaller component in its red wing, and corresponds to the pixel marked by a cross in the blueshifted region of the UV burst. Finally the black curve is a single-peaked profile centered at the rest wavelength of \ion{Si}{IV} 140.3~nm and belongs to the pixel marked by a cross in the white (zero velocity) region of the burst.

In the redshifted zones of Figure~\ref{fig_caso2_si4_vel} most of the pixels ($79\%$) have double-peaked profiles, while the remaining $21\%$ are single-peaked \ion{Si}{IV} profiles. These proportions change in the blueshifted zones of the UV burst, which are composed of $3\%$ with double-peaked profiles, $65\%$ with single-peaked profiles, and $32\%$ with single-peaked profiles that contain a lesser second component in one wing. Finally, the white areas of the velocity maps, i.e., where the \ion{Si}{IV} profiles are mainly at rest, are $100\%$ populated by single-peaked profiles. In this particular case the parts of the UV burst that are closer to disk center show a majority of double-peaked redshifted profiles, while the parts further away from disk center present mostly blueshifted (or at rest) single-peaked profiles.

Double-peaked \ion{Si}{IV} profiles are often interpreted as bidirectional jets expelled from the reconnection site both towards and away from the observer. Singled-peaked profiles with a smaller satellite component in either wing could be also interpreted as the observation of plasma ejected in one direction as a result of reconnection, plus another jet with a smaller Dopplershift ejected in the opposite direction in the line-of-sight. Overall, the structure of the observed velocities and intensities is suggestive of a nearly vertical current sheet, oriented in the $y$-direction, but slightly slanted with respect to the vertical. Reconnection will occur along such a current sheet as supported by Paper I, producing irregular bidirectional jets and, in this case, producing mainly downflows around $x=57\arcsec$ and mainly upflows around $x=58\arcsec$. The fact that profiles closer to disk center are double-peaked while those further away are mostly single-peaked with a satellite component is most likely due to projection effects. In parts of the burst pointing away from the observer where bidirectional jets are in place, one component may be damped respect to the other. 

\begin{figure}
\centering
\includegraphics[width=0.52\textwidth]{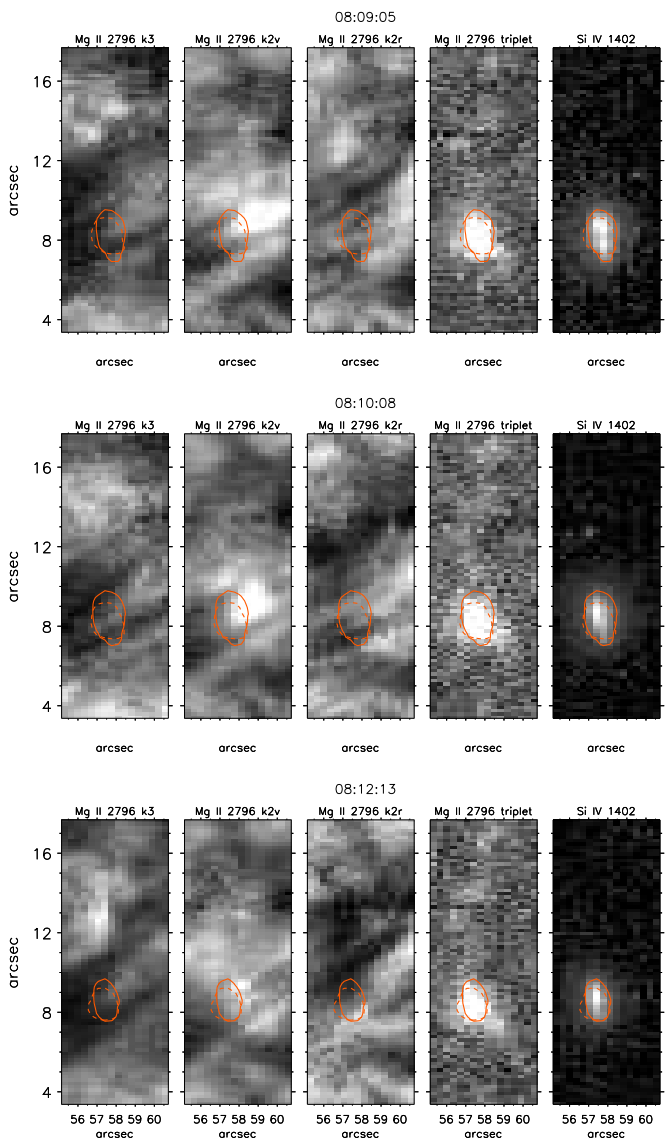}  
\caption{Raster intensity maps corresponding to the nominal positions of the \ion{Mg}{ii} k3, k2v, k2r spectral features, the \ion{Mg}{ii} triplet, and the \ion{Si}{iv}~140.3~nm line, for three timesteps around the maximum intensity peak of the UV burst. The solid orange contours delimit the UV burst as seen in the \ion{Si}{IV} intensity, while the dashed orange contours mark the boundaries of the burst as seen in the \MgII~triplet line feature. We have applied a gamma adjustment to the \ion{Si}{IV} color table.}
\label{fig_caso2_mg2_int}
\end{figure}

Figure~\ref{fig_caso2_mg2_int} shows intensity maps corresponding to IRIS rasters centered at the nominal wavelengths of the following line features (from left to right): \ion{Mg}{II} $279.6\,$nm k3, k2v, k2r, triplet and \ion{Si}{IV} 140.3~nm. The \ion{Si}{IV} maps show the intensity integrated over the whole line profile, unlike the \ion{Mg}{II} maps which are centered at a particular wavelength. We have selected the same timesteps around the maximum peak of intensity of the UV burst as in Figure~\ref{fig_caso2_si4_vel}. The solid orange contours delimit the UV burst as seen in the \ion{Si}{IV} intensity, while the dashed orange contours mark the boundaries of the burst as seen in the \MgII~triplet line feature. Note that the UV brightening is not visible at all wavelengths due to a canopy of overlying cold fibrils, therefore the contours have been derived from those wavelengths where the burst is always visible.  

Within the spectral rasters the morphology of the burst changes between the \ion{Mg}{II} and \ion{Si}{IV} maps: the UV brightening appears more slender in the FUV than in the NUV. This may be related to what is found in numerical experiments. In the simulations of Paper I (see e.g. Figure 1), the magnetic field expands with height so the reconnecting current sheet grows gradually wider, but in the direction perpendicular to the current sheet the visible width of the current sheet becomes thinner. This is a result of the plasma-$\beta$ (the ratio of gas pressure to magnetic pressure) decreasing with height: the magnetic field and the gas are more or less equally matched in the upper photosphere leading to rounder more amorphous structures there as the current sheet is deformed by plasma motions, but at greater heights the magnetic field completely dominates, the field lines straighten out and fill all space and the current sheet delineates two regions of equally strong straight field that are pointing in different (opposite) directions. Even within the same timestep of Figure~\ref{fig_caso2_mg2_int} the intensity maps in the different \ion{Mg}{II} features differ significantly from each other, as well as from the associated \ion{Si}{IV} map. In the k3 core we do not see any evidence of the UV burst at all at any time. A similar situation is observed in the \ion{Mg}{II} k2r peak, where only at 08:10:08 UT a slight hint of the UV brightening can be guessed. This is because diagonally oriented fibrils of cold material apparently cover the burst at both of these wavelengths. These cold fibrils are most likely related to the surge, as they follow the same diagonal direction and this is the period around which the surge reaches its maximum length and shows a very dynamical behavior. The situation changes in the k2v emission peak: at this wavelength we partially see a broader bright feature located in the same region as the \ion{Si}{iv} burst, but only for the first two rasters at 08:09:05 and 08:10:08 UT. This brightening is not visible in the last raster (only guessed) presumably because the aforementioned loops cover the brightening. Finally, the \MgII\ triplet 279.6~nm line shows the full extent of the UV burst, which as stated above is wider than what is seen in \ion{Si}{iv} maps. In fact, the UV burst in the \MgII\ triplet has a width of approximately $2 \arcsec$ (from x=56.5 to $58.5 \arcsec$), while in \ion{Si}{iv} is barely $1\arcsec$ wide (from $57\arcsec$ to $58\arcsec$).  At the same time the associated EB shows, in \halpha, an elongated shape during some timesteps alternating to a more roundish shape at other timesteps. Its size is smaller than the UV burst (around $1-1.5\arcsec$ of maximum length and $0.5-1\arcsec$ in width). This shape and size difference with rising temperature is explained by the expansion of the magnetic field with height; a few hundred kilometers above the photosphere where EBs are formed the magnetic field is confined by plasma motions in the high plasma-$\beta$ environment, but at greater heights the magnetic field expands and the current sheet that generates the EB and UV burst attains a wedge-like shape wider at the top than at the bottom (see Paper I) allowing for more extended brightenings. 

\begin{figure*}
\centering
\includegraphics[width=\textwidth]{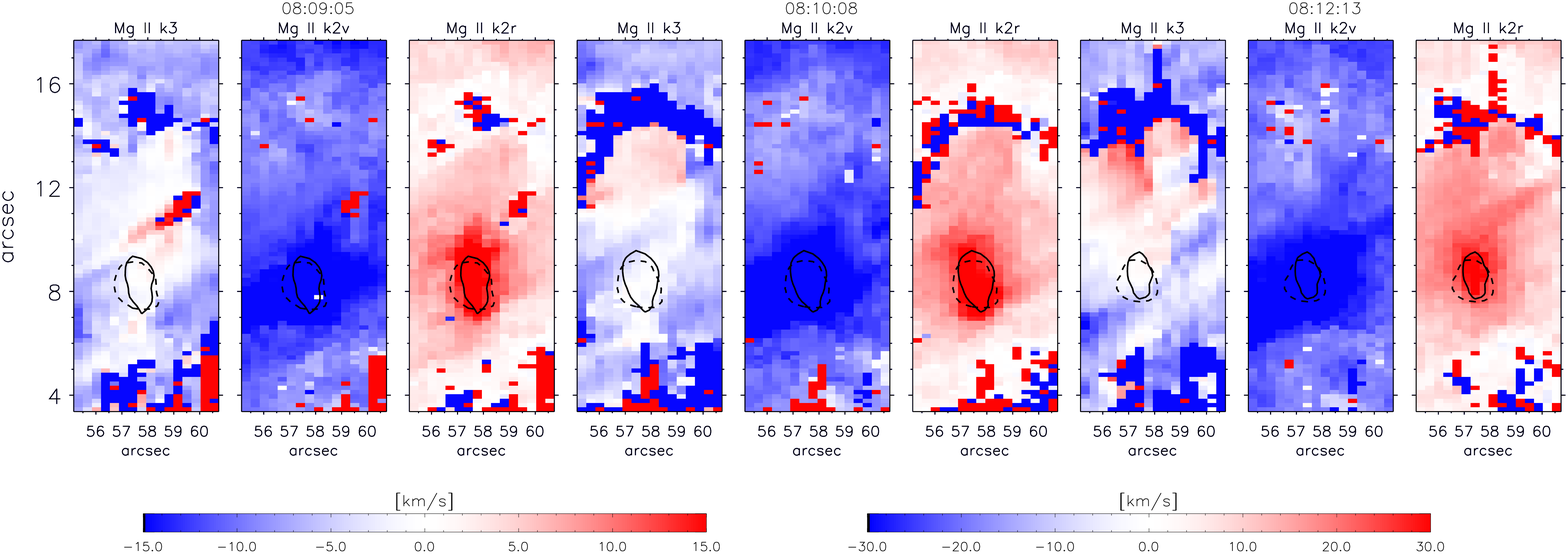}
\caption{Maps of the \MgIIk~Doppler shifts for three different timesteps. For each time there are three panels representing the shifts respect to the rest wavelength of the k3, k2v and k2r features of the \ion{Mg}{II}~k line. The color table on the left relates to the k3 spectral feature while the color table on the right refers to the k2v and k2r features. The solid black contours delimit the UV burst as seen in the \ion{Si}{IV} intensity, while the dashed black contours mark the boundaries of the burst as seen in the \MgII~triplet line feature.}
\label{fig_caso2_mg2_vel}
\end{figure*}

Finally Figure~\ref{fig_caso2_mg2_vel} presents Doppler shift maps of the \MgII~k3, k2v and k2r line features for the same three timesteps shown in Figures~\ref{fig_caso2_si4_vel} and \ref{fig_caso2_mg2_int}. These maps measure the observed departure of the particular spectral feature respect to their laboratory rest wavelength. In order to determine the positions of the maximums and minimums along the \ion{Mg}{II} k spectral line we have first smoothed the profiles both in the $y$-direction and in the $\lambda$-direction to avoid fluctuations that make it difficult for the fitting routine to find the proper k3, k2v and k2r spectral features. While most of the maps present smooth variations of the velocity values across the FOV, some pixels above and below the UV burst show extreme values with abrupt changes from positive to negative values. Those pixels correspond to \MgII~profiles in which the spectral features cannot be identified due to their atypical shape. The solid black contours delimit the UV burst as seen in the \ion{Si}{IV} intensity, while the dashed black contours mark the boundaries of the burst as seen in the \MgII~triplet line feature.

Unlike the \ion{Si}{IV} 140.3~nm Doppler shifts shown in Figure~\ref{fig_caso2_si4_vel}, the \ion{Mg}{II} Doppler shifts do not have differentiated zones within the UV brightening domain. In this case the \MgII~line profiles (like the ones shown in Figure~\ref{fig_caso2_spectra+timespectra}) have much broader wings than the profiles at rest. The k3 minimum presents almost no Dopplershift, the k2v maximum is blueshifted respect to the profiles at rest (with peak values of -40 \kms) and the k2r maximum is redshifted with respect to the profiles at rest (with maximum values of up to 40 \kms) for all timesteps. The shifts in k2v and k2r are not the result of monolithic mass motions, but due to other reasons. The entire region around the EB/UV burst phenomena is characterized by broad wings in the \MgII~k2 peaks, both on the blue side of the k2v peak and on the red side of the k2r peak. Spectral inversions of this line using the IRIS$^2$ package \citep[see {\tt http://iris.lmsal.com/iris2/} and][]{2019ApJ...875L..18S,2016ApJ...830L..30D} attribute the broad asymmetric \MgIIhk\ peaks most likely to the following factors in the region surrounding the EB/UV burst: a highly turbulent plasma, with turbulent velocities $v_{turb}$ of $10-15$~\kms, covering an area that is larger than the UV burst seen in the other plasma variables like temperature $T$, line of sight velocity $v_{los}$, and electron density $n_{\rm e}$. The maximum of $v_{turb}$ is reached in the lower chromosphere, where $\log(\tau_{500}) \simeq -3$ (where the photosphere is defined by the height where the optical depth $\tau = 1$ in the continuum at 500 nm, and thus the notation $\tau_{500}$). A region of smaller horizontal extent than what $v_{turb}$ covers, outlining the high intensities seen in figure~\ref{fig_caso2_mg2_int}, shows electron densities $n_{\rm e}\simeq 10^{13}$~cm$^{-3}$, which is an order of magnitude greater than the ambient densities. Filling the same extent, the inverted temperature rises to $7\,000$~K at $\log(\tau_{500})\simeq -3$, some $2\,000$~K hotter than ambient and remaining roughly constant up to $\log(\tau_{500})\simeq -6$. The velocities inferred from the inversions in the same vicinity are found to be upflowing and of order $-10$~\kms~to $-20$~\kms. This pattern fills the lower to middle chromosphere ($-2>\log(\tau_{500})>-4$). Overlying this turbulent, dense and relatively hot plasma IRIS$^2$ finds that the turbulent velocities and densities decrease while the upflow velocity gradually decreases towards zero. In the upper chromospheric emission ($\log(\tau_{500})<-6$) one sees downflowing plasma with a $u_{los}\simeq 10$~\kms~redshift that outlines the diagonal fibrils visible in \halpha~and the k3 and h3 cores. One possible interpretation of the Doppler shifts found in the k2v and k2r peaks is then that the reconnection associated with the EB and UV burst generates turbulent wave motions that propagate into the surrounding medium in a wider area than actually covered by the hot, dense plasma, while at the same time accelerating the plasma upwards and heating it in a narrower region around the current sheet. The inverted atmospheres found for the broad \MgII\ k2 and h2 peaks in the vicinity of the EB and UV burst yield very similar results for all the rasters we have examined.

The excellent temporal cadence of the IRIS observations allows us to discern the temporal evolution of the UV burst properties. During the 15 minutes that we observe the UV burst in our observations, 90 IRIS SJIs (both in $140\,$nm and in $279.6\,$nm) capture its evolution at a 10 s cadence. As mentioned before, the SJIs show a UV brightening with a predominantly circular shape that does not change much during its lifetime. In the spectral rasters the burst is smaller. The burst moves slightly towards disk center during the observations: a total of $-0.7\arcsec$ in the $x$-direction and $-1.4\arcsec$ in the $y$-direction. The \ion{Si}{IV} velocities reach peak values right before the maximum intensity at 08:10 UT and then decrease again, a behavior also seen in the \ion{Mg}{II} velocities.

\subsection{Coexistence of an EB and a UV burst: September 5, 2016}
\label{caso3}

\begin{figure*}
\centering
\includegraphics[width=0.85\textwidth]{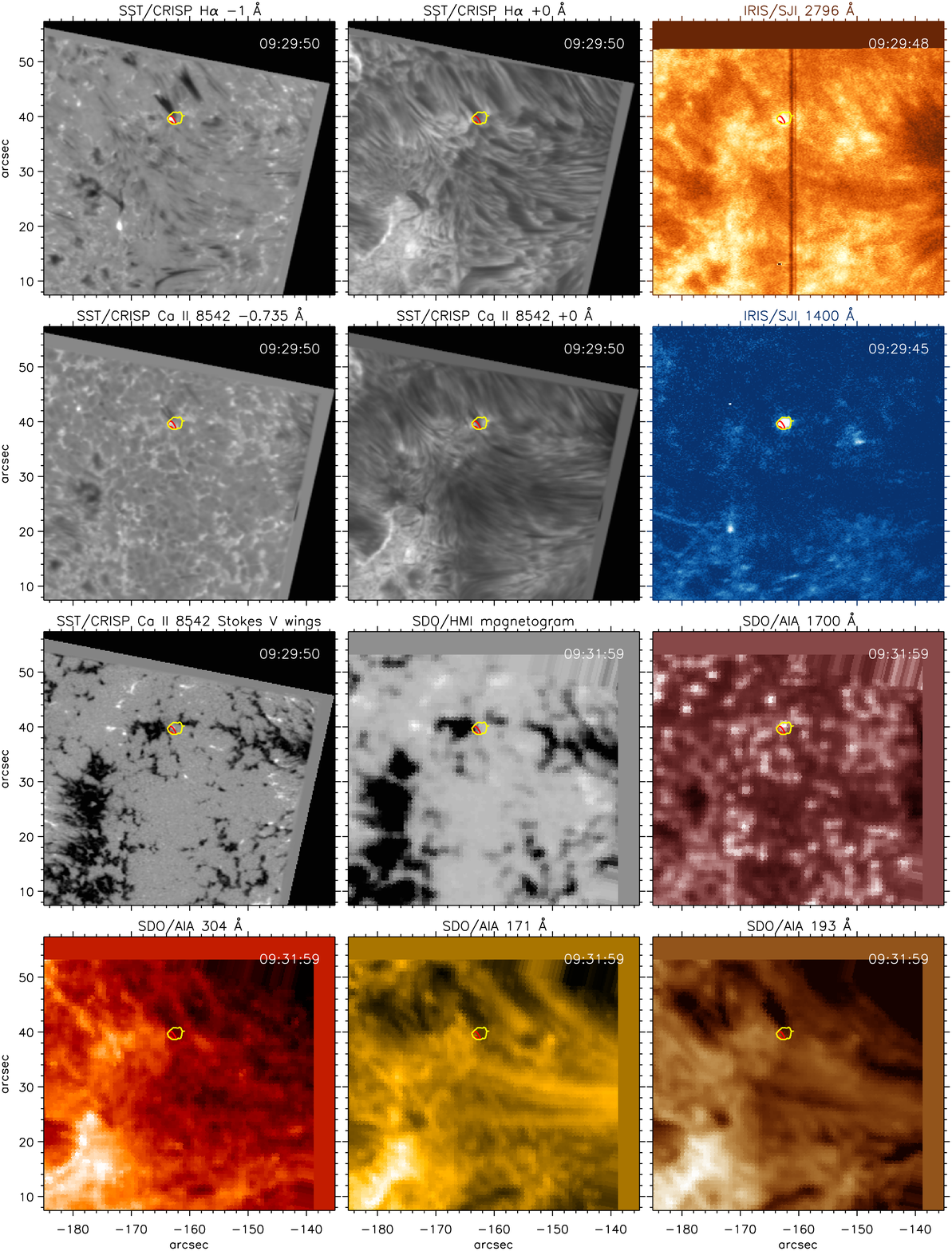}  
\caption{Same as Figure~\ref{fig_caso2_mos} for September 5, 2016. At this particular time both an EB and a UV burst co-existed co-temporally and co-spatially in the FOV.}
\label{fig_caso3_mos}
\end{figure*}

This case of transient brightenings on September 5, 2016 also showcases the coexistence of an EB and a UV burst both spatially and temporally. The coordinates of these transient events are as follows: the initial position of the EB was $(x,y)=(-162.5,39.5) \arcsec$ starting at 09:26:04 UT and the initial position of the UV burst was $(x,y)=(-162.5,39.8) \arcsec$ at 09:26:36 UT. Figure~\ref{fig_caso3_mos} presents the mosaic of SST, IRIS and SDO observations that shows the context in which these brightenings were produced. The SST images in Figure~\ref{fig_caso3_mos} were taken at 09:29:50 UT, the IRIS SJI $140\,$nm at 09:29:45 UT and the SJI $279.6\,$nm at 09:29:48 UT. Finally, the SDO imagery was taken at 09:31:59 UT. A flux emergence region is situated in the center of the FOV, East of the penumbra seen in the right side of the panels and North-West of the small penumbrae observed on the left side of the panels. As in Figure~\ref{fig_caso2_mos}, the EB is marked with a red contour and the UV burst with a yellow contour. The location where these brightenings occur coincides with fields of opposite polarities colliding with each other as can be seen in the Stokes V maps of \ion{Ca}{II}~854.2~nm or the HMI magnetograms of Figure~\ref{fig_caso3_mos}. An extended patch of negative polarity interacts with a very small patch of positive polarity. 

In the SST images the EB reaches a maximum size of around $1 \arcsec \times 1.5 \arcsec$ with a very varying and irregular shape along its lifetime. The UV burst, who in the SJIs lies always on top of the EB, maintains a roundish shape during its lifetime and peaks at a diameter of $2.5 \arcsec$. The burst is observed both in the NUV SJI at $279.6\,$nm as well as in the FUV SJI at $140\,$nm. 

As in the previous case (and in all cases where a UV burst is present, see Table~\ref{table1}) the \halpha\ wing images at $-0.1$~nm~ show a surge shooting diagonally from the EB in the North-East direction at 09:27:09 UT and lasting for 6 minutes. The maximum length of the surge is $9 \arcsec$ in the \halpha~$-0.1$~nm images and $12 \arcsec$ in the \halpha~core images. The temporal evolution of the surge can be followed in the \halpha~and \ion{Ca}{II}~854.2~nm movies presented as supplemental material. 

The response of the AIA transition region and coronal channels is very similar to that of the case introduced in Section~\ref{caso2}, i.e., no signature associated to these reconnection events is observed at those hotter EUV wavelengths. The surge, on the other hand, leaves a visible imprint in the form of a darkening with the same shape. Again the cold dense material of the surge absorbs light and does not allow radiation escaping from the reconnection site to reach coronal heights. However, at much lower chromospheric temperatures, the AIA 170~nm channel displays a small brightening at the co-spatial position of the EB and the UV burst. 

\begin{figure}
\centering
\includegraphics[width=0.50\textwidth]{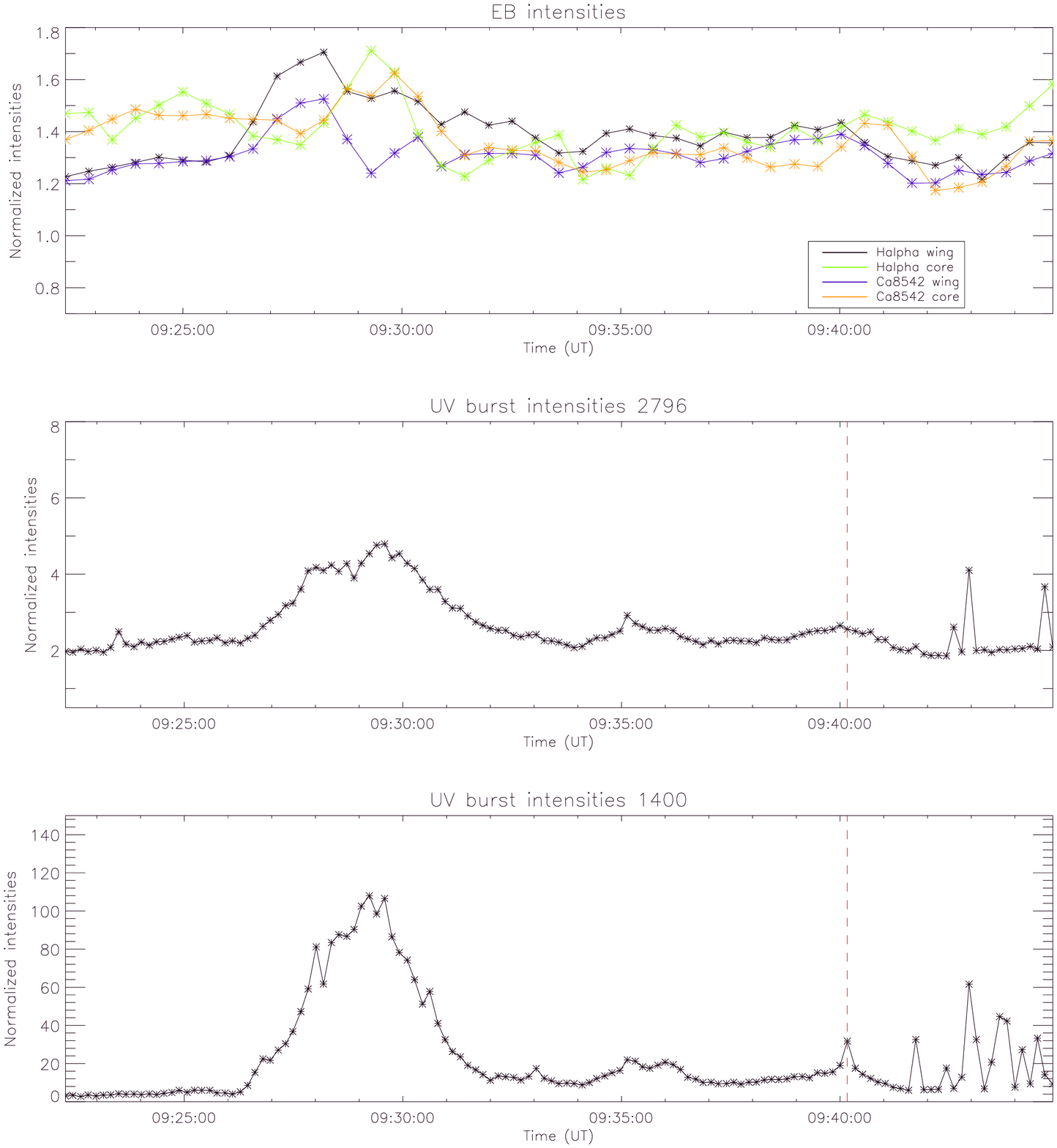} 
\caption{Same as Figure~\ref{fig_caso2_evtemp} for September 5, 2016. The red vertical dashed line marks the start of the IRIS satellite's passage through the South Atlantic Anomaly where the detectors encounter increased noise levels due to cosmic ray hits.}
\label{fig_caso3_evtemp}
\end{figure}

Figure~\ref{fig_caso3_evtemp} illustrates the temporal evolution of the EB and UV brightening intensities during their coexistence. As in Figure~\ref{fig_caso2_evtemp}, the upper panel displays the intensities of the EB in \halpha~(wing and line core) and in \ion{Ca}{II}~854.2~nm (wing and line core). The middle panel shows the intensity derived from the IRIS SJIs $279.6\,$nm while the lower panel presents the intensity obtained from the SJI $140\,$nm images (see details on the identification and tracking of events and how light-curves are calculated in Sects.~\ref{EB_iden} and \ref{UV_iden}). Note that in this case the IRIS observations are affected by SAA events at the end of the observing period and those can be seen as outlier peaks in both UV intensities. The beginning of the SAA period is marked by red dashed vertical lines in Figure~\ref{fig_caso3_evtemp}.

The EB is present during almost the whole observation period, but it shows a peak in intensity between 09:26:04 until 09:33:35 UT (with a maximum value at 09:28:13 UT). At the end of the observing period the EB is still visible on the solar disk. Almost simultaneously (only 30 s later) the burst in NUV and FUV lights up at 09:26:36 UT lasting for 7.5 minutes. At the end of the observation the UV burst is still slightly visible in the SJI $140\,$nm. The maximum intensity is observed between 09:29:14 and 09:29:35 UT in \ion{Si}{IV} (lower panel) and at 09:29:35 UT in \ion{Mg}{II} (middle panel), that is, 80 s after the maximum in \halpha~wing and \ion{Ca}{II}~854.2~nm wing. However, note that the line core of \halpha~peaks at the same time as the UV intensities. 

\begin{figure*}
\centering
\resizebox{0.49\hsize}{!}{\includegraphics{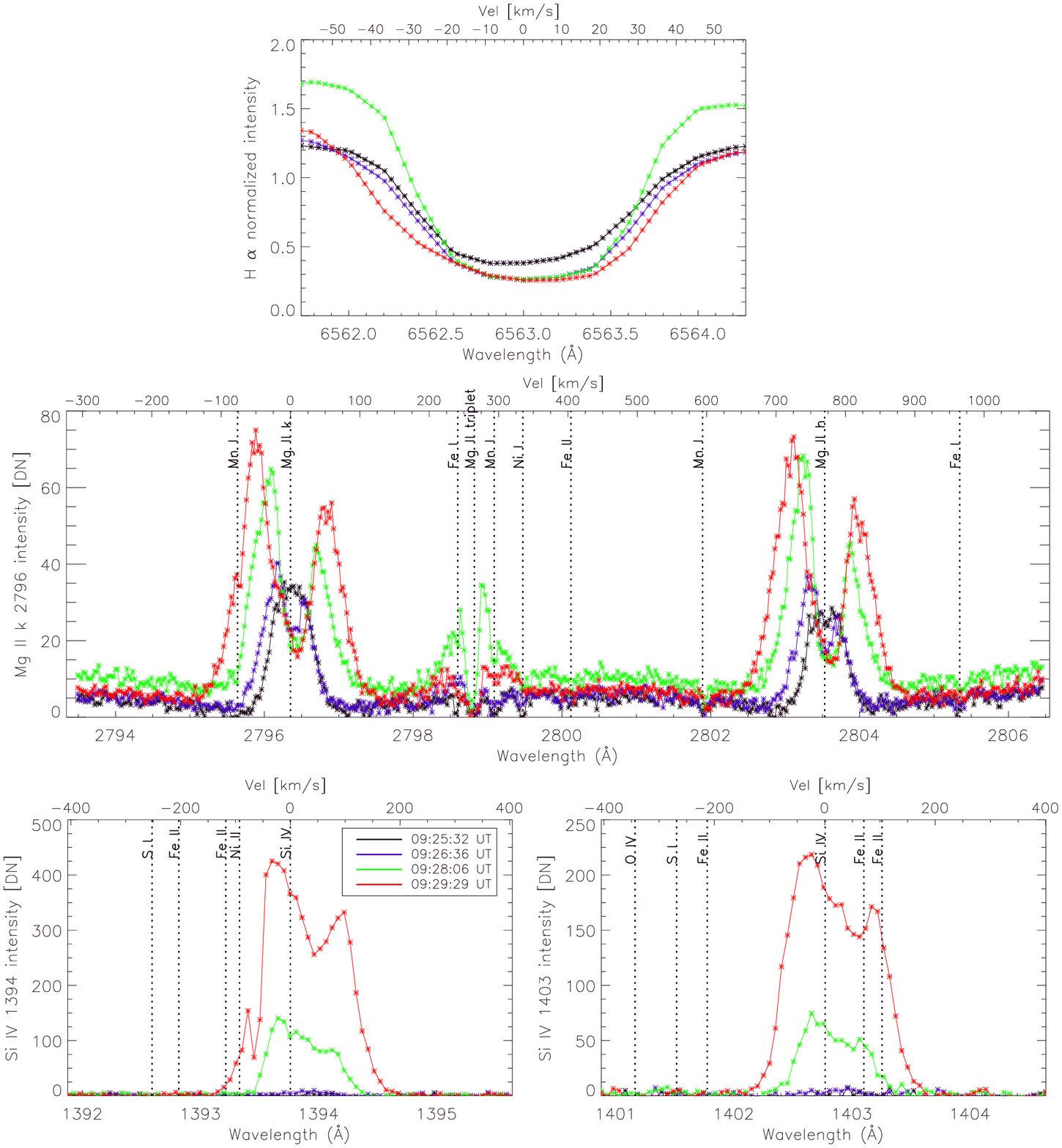}}
\resizebox{0.49\hsize}{!}{\includegraphics{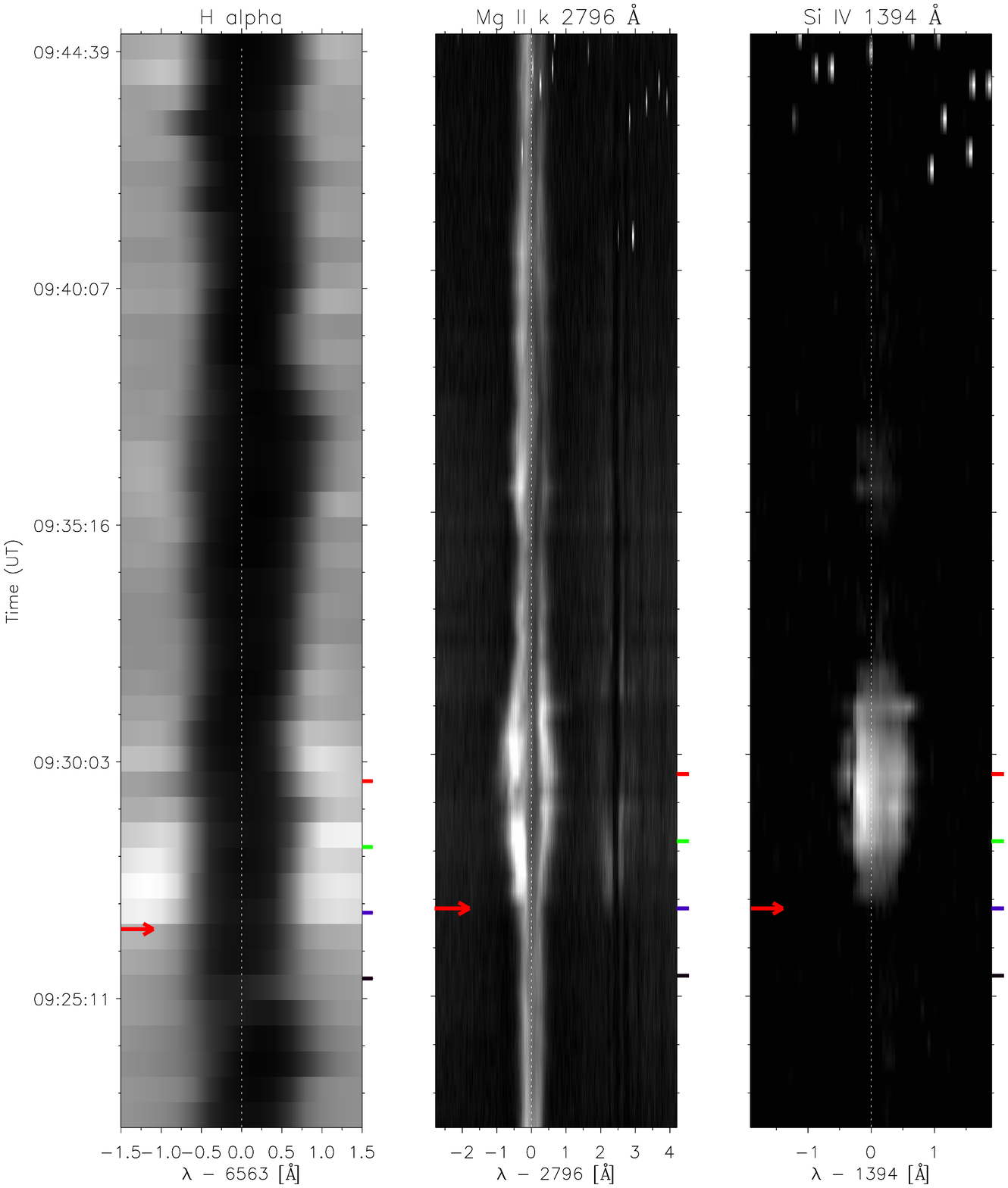}}
\caption{Same as Figure~\ref{fig_caso2_spectra+timespectra} for September 5, 2016. The colored marks in the time-sliced spectra pinpoint to the same times that have been highlighted in the left panel. We have applied a gamma adjustment to the \ion{Si}{IV} color table.}
\label{fig_caso3_spectra+timespectra}
\end{figure*}

To complete this case, we present in Figure~\ref{fig_caso3_spectra+timespectra} the spectral information gathered from the SST and IRIS observed spectra. The left panel shows, like in Figure~\ref{fig_caso2_spectra+timespectra}, spectral profiles for \halpha~(top panel), \ion{Mg}{II}~k and h lines at $279.6\,$~nm (middle panel), and the \ion{Si}{IV} lines at 139.4 and 140.3~nm respectively (bottom panels), for four selected time steps. These times represent, approximately: the onset of the EB (09:25:32 UT, black line), the onset of the UV burst (09:26:36 UT, blue line), maximum brightness of the EB (09:28:06 UT, green line), and finally maximum brightness of the UV burst (09:29:29 UT, red line). The right side of the figure is a time-sliced spectra that displays the temporal evolution of three spectral lines: SST \halpha, IRIS \ion{Mg}{II}~k and triplet lines and IRIS \ion{Si}{IV} 139.4~nm. The red arrow in the \halpha~spectra highlights the onset of the EB, and the red arrows in the IRIS spectra mark the onset of the UV burst. 

The upper left panel shows that the \halpha~spectra are quite similar at all times, with the exception of 09:28:06 UT (green curve, moment of maximum brightness of the EB), when the wings of \halpha~exceed 150\% of the intensity of a quiet Sun profile. In fact, the left wing reaches 170\% of that intensity. It is indeed that same timestep (09:28:06 UT) who presents the highest emission in the \MgII~triplet in the central left panel of Figure~\ref{fig_caso3_spectra+timespectra}. Both the \ion{Mg}{II} spectra taken at 09:28:06 and 09:29:29 UT show a very slight absorption due to \ion{Mn}{I} superimposed in the blue wing of \ion{Mg}{II} k profiles. This absorption is not visible in the other two timesteps (who coincide with the onset of the EB and UV burst) as their profiles are not broad enough to present absorption lines. An inspection to the \ion{Si}{IV} spectra in the lower left panels reveals that only these same timesteps (09:28:06 and 09:29:29 UT) have a significant amplitude in the FUV. Only the 09:29:29 UT profile is wide enough to present \ion{Ni}{II} absorption lines at $-93$~\kms~superimposed in the \ion{Si}{IV} 139.4 nm profiles.

This scenario suggests that in the beginning of the observation the atmosphere is not hot enough. It is only around 09:27 UT and beyond that the lower chromosphere is hot enough to produce emission in the triplet of \ion{Mg}{II} and later on heats even more to produce a significant intensity amplitude in the FUV \ion{Si}{IV} profiles. This sequential behavior is also seen in the time-sliced spectra (right side of Figure~\ref{fig_caso3_spectra+timespectra}): first the EB lits up maintaining its bright wings for about 2.5 minutes (the right wing is bright for a longer time), then the NUV lits up 30 s after the EB does, and then the FUV starts heating up gradually reaching a significant intensity from 09:28 UT onwards. 

\subsection{EB not associated with a UV burst: September 5, 2016}
\label{caso4}

The event introduced in this Section presents a case of an EB not associated to any UV brightening, neither temporally nor spatially. This EB was present on the solar disk on September 5, 2016. Its initial position was $(x,y)=(-158.0,3.5) \arcsec$ at 08:16:36 UT. This is the only case of this study that does not have a surge associated with it. The EB is the site of interaction between magnetic fields of opposite polarities. In particular an extended and still negative polarity patch intermittently interacts with tiny positive polarity patches that constantly emerge at the edge of the negative patch. 

The EB of this case reaches a maximum size of $1 \arcsec \times 1 \arcsec$, varying along its life time. During the 16.5 minutes of life its size peaks at 08:19 and 08:22 UT. Its shape is irregular and also changes with time. No signature of this event is observed in the IRIS SJIs nor spectra. 

\begin{figure}
\centering
\resizebox{\hsize}{!}{\includegraphics{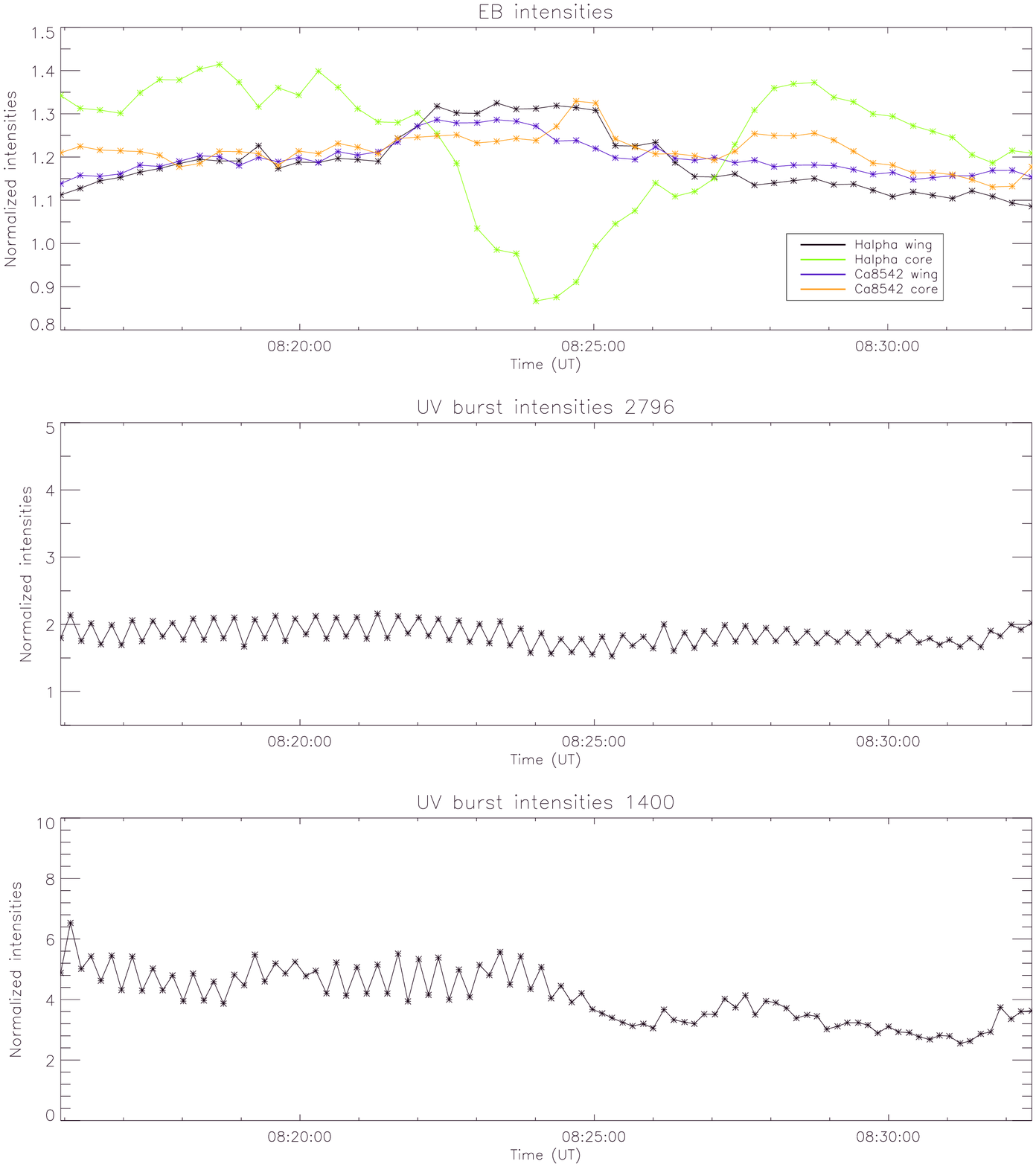}}
\caption{Same as Figure~\ref{fig_caso2_evtemp} for September 5, 2016. In this case only an EB was present and no UV brightening could be associated with it.}
\label{fig_caso4_evtemp}
\end{figure}

Figure~\ref{fig_caso4_evtemp} presents the temporal evolution of the intensity of this EB. The outlay of the figure is the same as in Figures~\ref{fig_caso2_evtemp} and \ref{fig_caso3_evtemp}. Intensity in the \halpha~and \ion{Ca}{II}~854.2~nm wing narrowband filtergrams is very similar and does not reach very high values compared to the previous case. However, the \halpha~line core (green curve) shows a dip in intensity around the time of maximum brightness of the EB (as measured by the wings, at 08:23 UT) which is not observed in the center of the \ion{Ca}{II}~854.2~nm line. From 08:22:20 to 08:27:24 UT an unrelated surge (whose feet lie around $4 \arcsec$ diagonally up-right from our event) covers the EB in its entirety causing a decrease in the \halpha~core intensity. The presence of that surge and its evolution can be seen in the accompanying supplemental materials. The EB starts at around 08:17 and ends at 08:26 UT lasting thus for 9 minutes. The UV intensity obtained from the IRIS SJIs (middle and lower panels) show no trace of the existence of any brightening. 

\begin{figure*}
\centering
\resizebox{0.49\hsize}{!}{\includegraphics{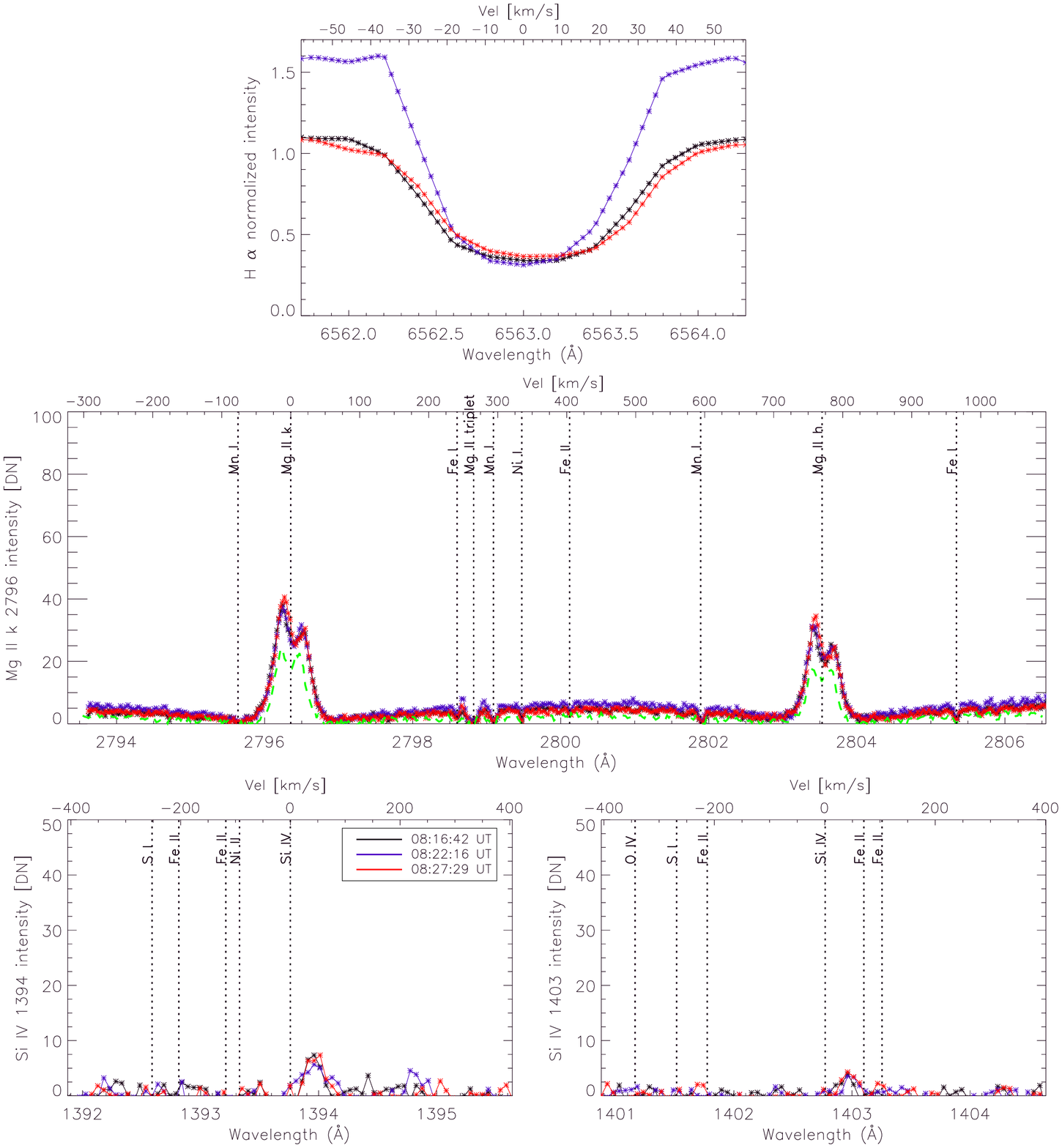}}
\resizebox{0.49\hsize}{!}{\includegraphics{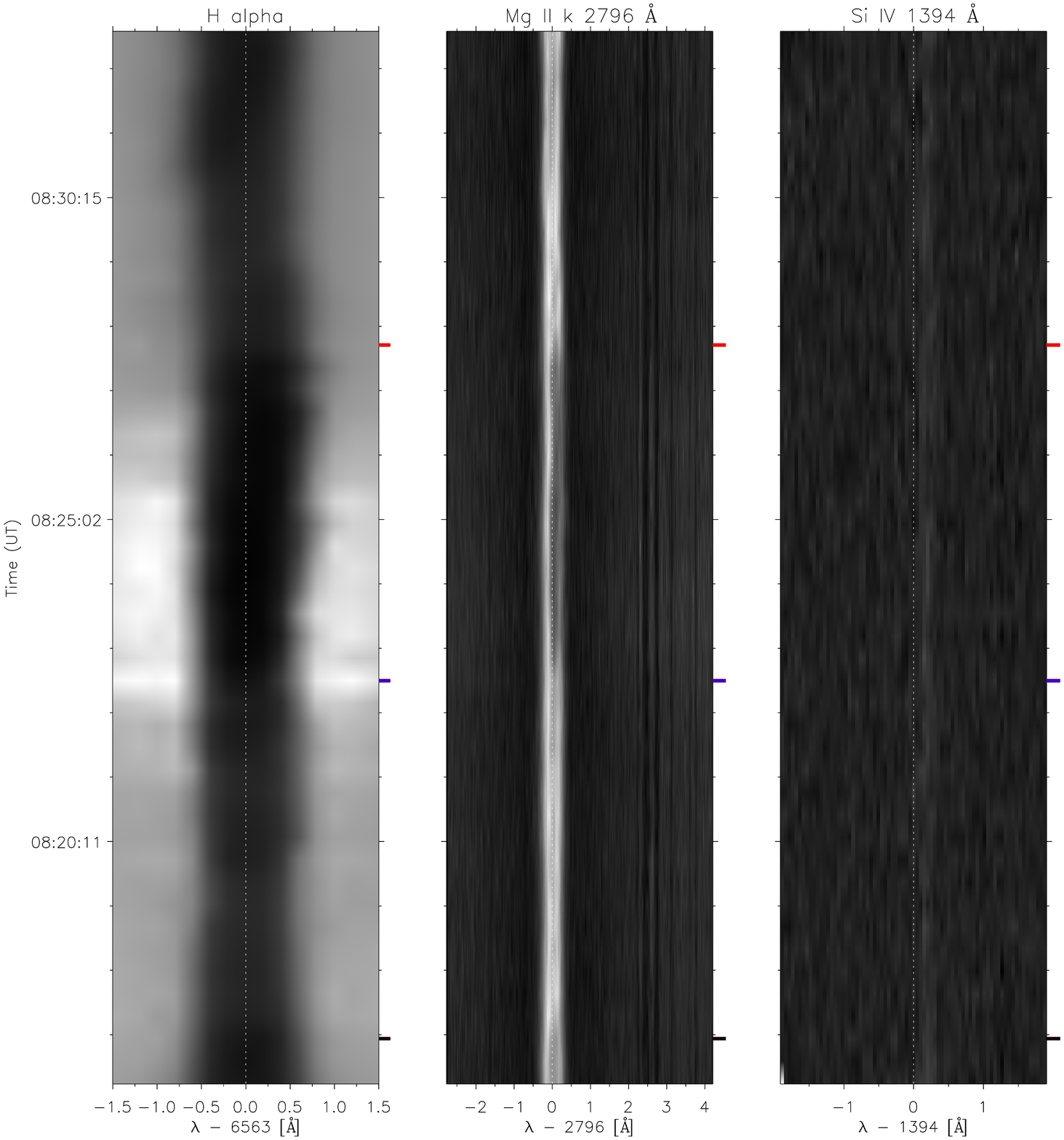}}
\caption{Same as Figure~\ref{fig_caso2_spectra+timespectra} for September 5, 2016. No UV burst was found associated with this particular EB. The dashed green curve in the \MgII~panel shows a quiet Sun profile for reference: barely any imprint from the EB is observed in the \MgII~spectra. The colored marks in the time-sliced spectra pinpoint to the same times that have been highlighted in the left panel.}
\label{fig_caso4_spectra+timespectra}
\end{figure*}

Figure~\ref{fig_caso4_spectra+timespectra} gives the spectral information on this event obtained from SST and IRIS at three different timesteps: beginning of the EB (black curve, 08:16:42 UT), approximate moment of maximum brightness of the EB (blue curve, 08:22:16 UT), and decay of the EB (red curve, 08:27:29 UT). The outlay of the figure is the same as in Figure~\ref{fig_caso2_spectra+timespectra} and Figure~\ref{fig_caso3_spectra+timespectra}. In the period of peak brightness the wings of \halpha~climb up to 160\% compared to the intensity of a quiet Sun profile. Both wings appear very symmetric. The profiles from the beginning and end of the EB's life (black and red curves) are only 110\% brighter than that of a quiet Sun pixel. Both the NUV and FUV profiles in the middle and lower panels show no significant brightening. Only a marginal intensity increase is observed in the \MgII~triplet at a particular timestep. 

The time-sliced spectra in the right side of Figure~\ref{fig_caso4_spectra+timespectra} shows only the typical spectral evolution of an EB whose peak intensity lasts for about three min. No apparent heating of the upper atmospheric layers can be measured in this case.

\subsection{UV burst not associated with an EB: September 5, 2016}
\label{caso7}

\begin{figure*}
\centering
\resizebox{0.85\hsize}{!}{\includegraphics{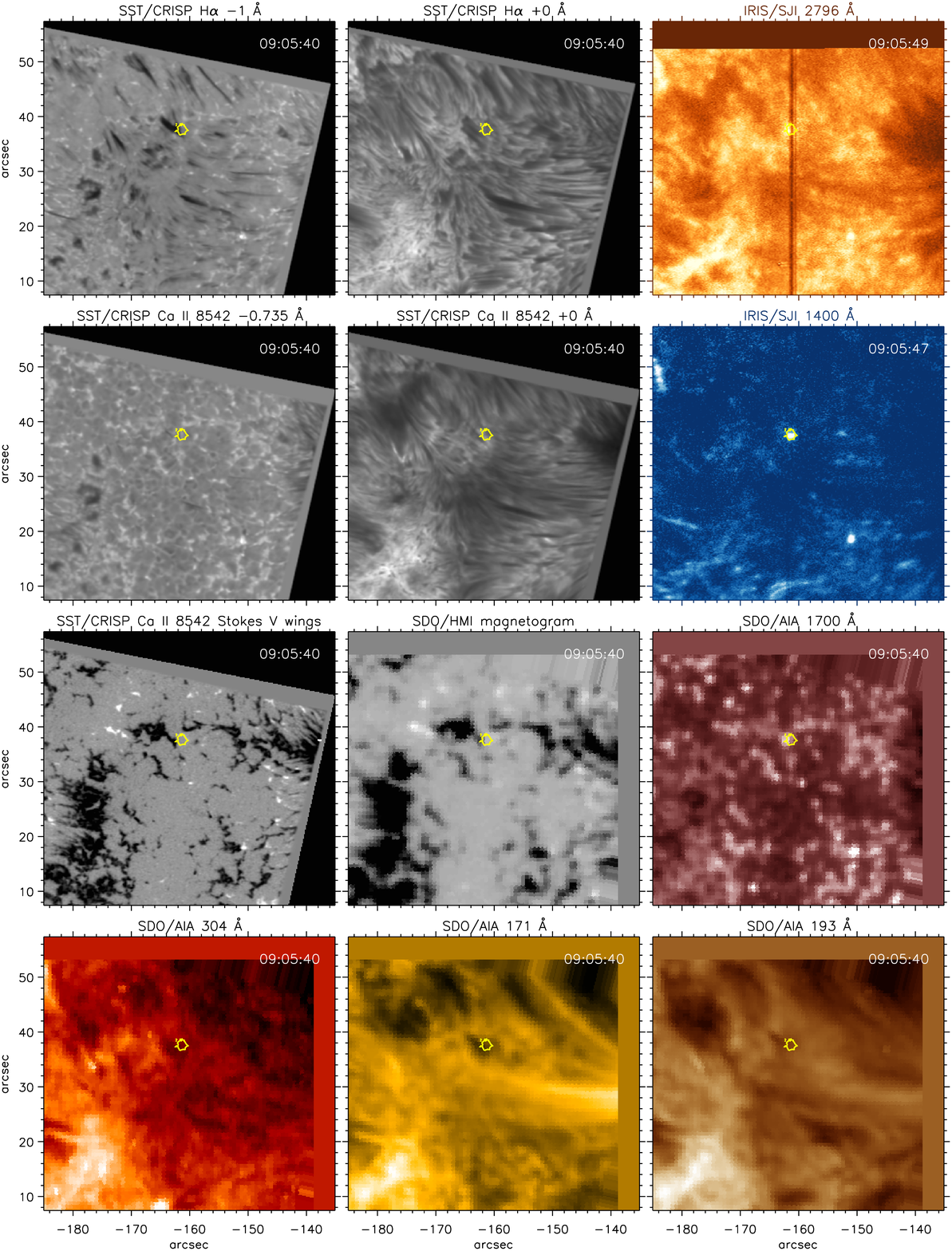}}
\caption{Same as Figure~\ref{fig_caso2_mos} for September 5, 2016. In this FOV and at this time of the observation only a UV burst was present with no associated EB.}
\label{fig_caso7_mos}
\end{figure*}

\begin{figure}
\centering
\resizebox{\hsize}{!}{\includegraphics{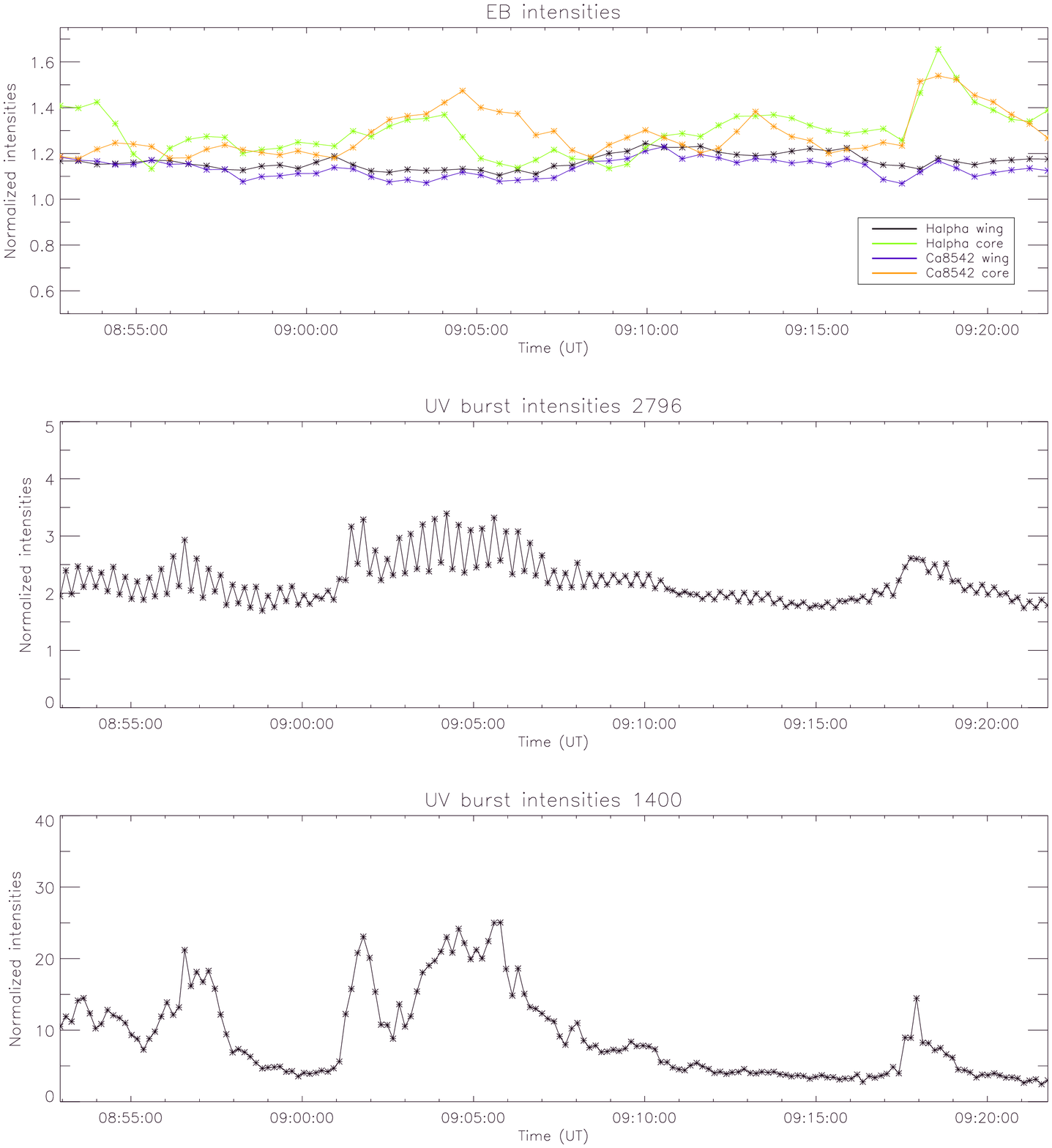}}
\caption{Same as Figure~\ref{fig_caso2_evtemp} for September 5, 2016. At this time of the observation only a UV brightening was present with no EB associated with it.}
\label{fig_caso7_evtemp}
\end{figure}

The last case we will show in detail is an event where only a UV burst was observed and no EB was found co-temporally nor co-spatially. On September 5, 2016 a UV brightening was observed in the solar disk from 08:53 until 09:19 UT. The brightening was indeed already visible at the start of the dataset. Its initial position was $(x,y)=(-160.7,37.1) \arcsec$ at 08:53:00 UT. This event is introduced in Figure~\ref{fig_caso7_mos} which sets the scenario surrounding this event. The FOV is the same as in Sect.~\ref{caso3} but at an earlier time, i.e., an area of active flux emergence between two sunspots. The Stokes V movies in the wings of \ion{Ca}{II}~854.2~nm show several small positive polarity patches traveling eastwards away from the sunspot in the upper right of the images and interact with the more extended negative polarity patch which lies at around $(x,y)=(-166,40) \arcsec$. In Figure~\ref{fig_caso7_mos} the displayed SST and SDO observations are taken at 09:05:40 UT, while the IRIS panels are taken at 09:05:47 and 09:05:49 UT for the \ion{Si}{IV} and \ion{Mg}{II} SJIs respectively. In this figure the UV burst is highlighted with a yellow contour that has been superimposed on all panels of the figure. The UV brightening shows a peak size of $3 \arcsec \times 2 \arcsec$ at 08:57:36 and at 09:03:31 UT. The brightening has a very bursty behavior (see details in Sect.~\ref{UVburst}), reaching several maximums of intensity and size as can be seen in the IRIS SJI movies, as well as in Figure~\ref{fig_caso7_evtemp}.

The \halpha~wing panel at $-0.1$ nm shows a surge shooting out diagonally from the lower-left part of the burst, which stays for the whole dataset. The surge is also visible in the core of the \halpha~line. Like in the previous cases, the surge is a very dynamical structure with a maximum length of 7$\arcsec$ (reached at 09:09:25 UT) in the \halpha~$-0.1$~nm images.

The SDO/AIA coronal channels do not show any signature of heating, but a darkening in coincidence with the surge. The AIA 170~nm panel of Figure~\ref{fig_caso7_mos} shows a brightening at the position of the UV patch.

Figure~\ref{fig_caso7_evtemp} shows the temporal evolution of the intensity at several wavelengths for the duration of the UV burst. Outlay is similar to that of Figures~\ref{fig_caso2_evtemp}, ~\ref{fig_caso3_evtemp} and ~\ref{fig_caso4_evtemp}. The wings of \halpha~and \ion{Ca}{II}~854.2~nm (black and blue curves) present a flat behavior (typical for a case with no EB), while the centers of those lines show some peaks in coincidence with brightenings occurring at the feet of the surge. The NUV and FUV intensities (middle and lower panels) obtained from the IRIS SJIs evolve in unison and present a bursty behavior with up to four peaks during the span of the UV burst total lifetime (about 26 min). During that time span, we can distinguish three bright periods of 6, 7 and 1 min of duration respectively in the \ion{Si}{IV} intensity. The intensity variations at these wavelengths are, for this case, much lower when compared to the bursts of Sects.~\ref{caso2} and ~\ref{caso3} when the UV bursts where associated with an EB. 

\begin{figure*}
\centering
\resizebox{0.49\hsize}{!}{\includegraphics{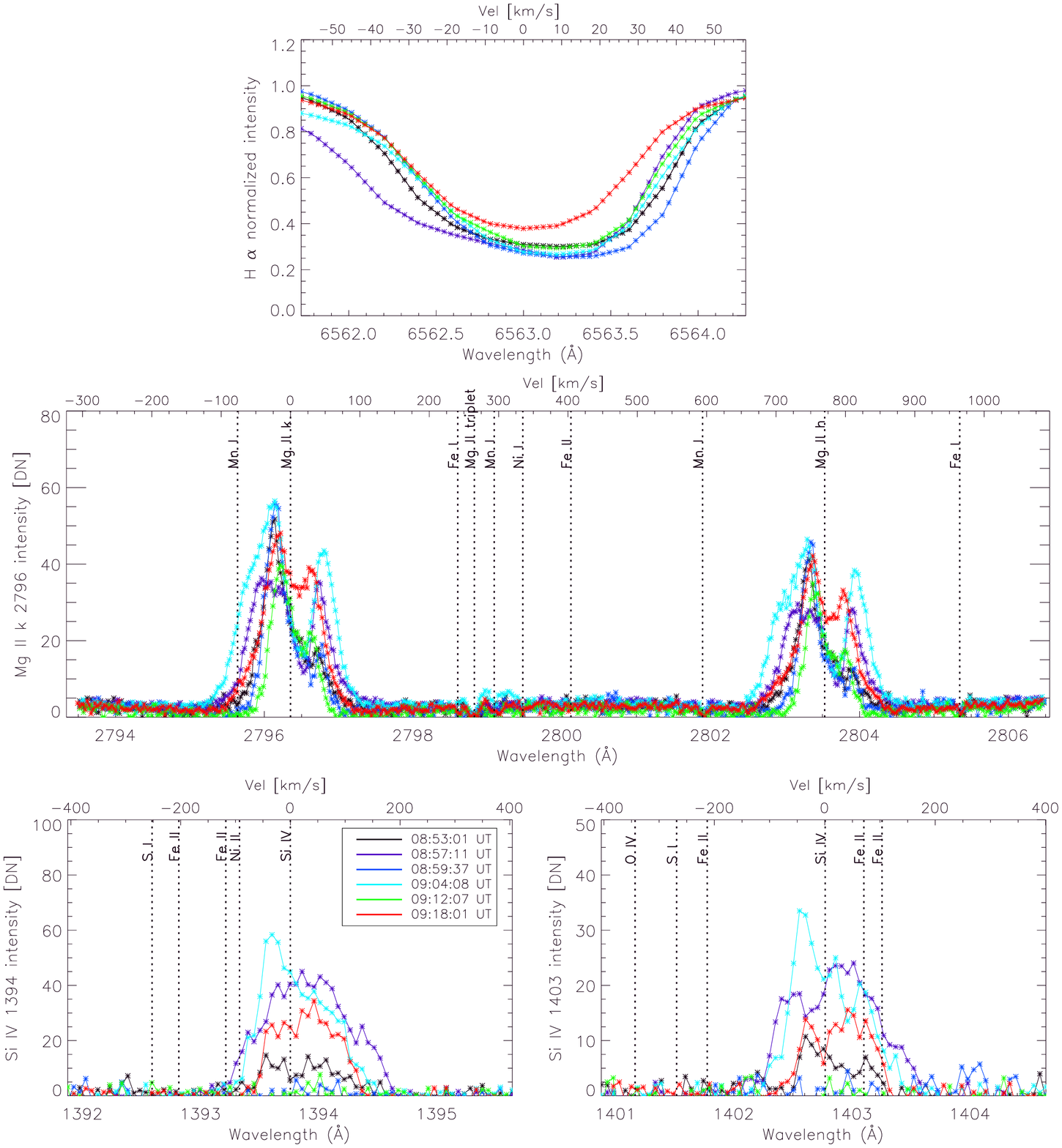}}
\resizebox{0.49\hsize}{!}{\includegraphics{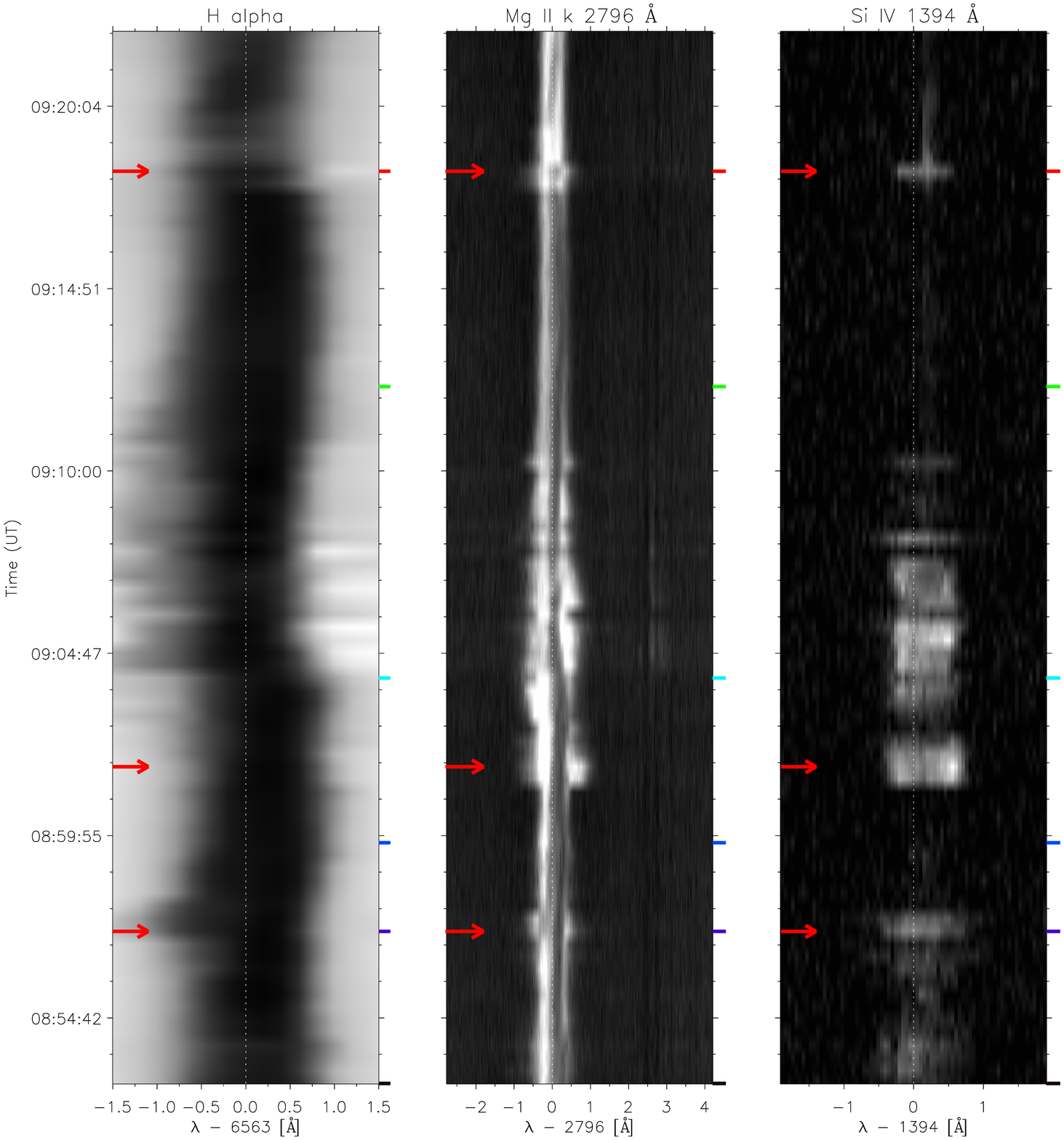}}
\caption{Same as Figure~\ref{fig_caso2_spectra+timespectra} for September 5, 2016 for a case where only a UV burst was present in the FOV. No EB was found that could be related to it. The colored marks in the time-sliced spectra pinpoint to the same times that have been highlighted in the left panel. We have applied a gamma adjustment to the \ion{Si}{IV} color table.}
\label{fig_caso7_spectra+timespectra}
\end{figure*}

Finally, Figure~\ref{fig_caso7_spectra+timespectra} shows the spectral profiles as well as their temporal evolution. The outlay is the same as in Figures~\ref{fig_caso2_spectra+timespectra}, ~\ref{fig_caso3_spectra+timespectra} and ~\ref{fig_caso4_spectra+timespectra}. The left side of the figure shows the SST and IRIS spectra at six different times: beginning of the observations (black line, 08:53:01 UT), first peak of the UV burst (purple line, 08:57:11 UT), relaxation time between peaks (dark blue, 08:59:37 UT), second peak of UV burst (turquoise line, 09:04:08 UT), relaxation time between peaks (green line, 09:12:07 UT) and finally third peak of burst (red line, 09:18:01 UT). The \halpha~profiles (upper left panel) do not show any particular feature at any time, other than a blueshift at 08:57:11 UT and a redshift at 08:59:37 UT (both Dopplershifts visible also in the time-sliced spectra). The UV spectra in the middle and lower left panels show the typical signature of a UV burst but at a much smaller scale than in the previously presented cases. No photospheric or chromospheric absorption lines are found superimposed to either the \ion{Mg}{II}~k and h profiles at $279.6\,$~nm nor the \ion{Si}{IV} profiles at 139.4 and 140.3~nm. One can note that only the moments of peak intensity (purple, turquoise and red curves) have a measurable amplitude in the \ion{Si}{IV} profiles, while the periods of relaxation between peaks do not show a significant FUV profile. 

The time-sliced spectra on the right side of Figure~\ref{fig_caso7_spectra+timespectra} displays the temporal evolution of three spectral lines: SST \halpha, IRIS \ion{Mg}{II}~k and triplet lines and IRIS \ion{Si}{IV} 139.4~nm. The red arrows mark the beginning of the different bright periods along the 26 min of lifetime. The \halpha~temporal evolution shows two periods of blueshifts lasting for about one and eight min respectively. Those blueshifts roughly correspond to the periods of brightening in the UV wavelengths.

\subsection{Other cases in brief}

In addition to the four cases introduced in more detail in the previous Sections we have also examined four more cases that we briefly report here. Two of these cases are events in which an EB and a UV burst coexisted co-spatially and co-temporally, both of them occurring on September 3, 2016 at two different locations within the same FOV. In these particular cases the EB and the UV burst lit up almost simultaneously (within one min). On September 4th a case of a UV burst not associated to an EB was identified away from the IRIS slits. Finally our last case was on September 5th also showing a UV burst not accompanied by an EB. We believe that this UV brightening falls in the FAF (flaring arch filament) category \citep{2015ApJ...812...11V}. All these UV brightenings share an intermittent (bursty) behavior when looking at the temporal evolution of their intensities and the supplemental movies. In addition, all these four cases have surges associated to them. More details can be found in Table~\ref{table1}. 

Finally we have derived the velocities of our seven UV brightenings using the observed IRIS lines. The \ion{Si}{IV} 140.3~nm velocities yield maximum values of $\pm 100$ \kms~for all cases, with averages around $\pm 50$ \kms. Regarding velocities derived from the IRIS \ion{Mg}{II} lines, average values for the \MgII~k3 minimum yield $\pm 5$ \kms. The velocities for the k2v and k2r maximums reach $\pm 30-40$ \kms~in all cases. 

\subsection{Absence or presence of \ion{O}{IV} lines}

\citet{2014Sci...346C.315P} realized that the forbidden \ion{O}{IV} 139.97 and 140.1 nm lines were absent in their identified UV bursts. \citet{2016ApJ...824...96T} noted that those lines were also absent or very weak in their UV brightenings associated with EBs, but present in the bursts not connected to EBs. 

The existence of \ion{O}{IV} lines and the ratio of \ion{Si}{IV} 139.4 nm to \ion{O}{IV} 140.1 nm provides a lower limit of the electron density in the source region of these lines \citep[see, e.g.][]{2014Sci...346C.315P}. These authors used this method to conclude that the most likely scenario for the absence of \ion{O}{IV} lines in their observations of UV bursts is due to high density in the bombs, reaching $3\times10^{14}$ cm$^{-3}$ in the source region of Mg II, implying formation very low in the solar atmosphere. Other justifications for this absence have been given: \citet{2013ApJ...767...43O} attribute the absence of \ion{O}{IV} lines to non-equilibrium ionization and \citet{2014ApJ...780L..12D} states that, even at low densities, such lines can be greatly suppressed in the presence of non-Maxwellian electron distributions.

Our IRIS observations do not allow us to conclude anything in this regard because the exposure time of the IRIS spectral observations was only 0.5 s, and those lines are faint and require longer exposure times (for reference, \citet{2016ApJ...824...96T} had a exposure time of 8 s), resulting in profiles that are too noisy for an accurate detection of any \ion{O}{IV} signal above the noise level.

\section{Discussion and Conclusions}
\label{discu}

AR 12585, a region of abundant flux emergence which we observed in September 2016, has served the purpose of studying the temporal and spatial connection between transient dynamical phenomena resulting from magnetic reconnection, namely EBs and UV bursts. For this purpose we took, reduced and analyzed high cadence, high resolution coordinated observations from the SST and IRIS. The eight cases studied allow us to establish a clear relationship between them, which we interpret in view of the results of the numerical Paper I. 

We find that EBs are sometimes co-spatial and co-temporal with a UV burst, and sometimes not. The same is true for UV bursts, which are sometimes colocated and simultaneous with an EB, and sometimes not. In a given spectral image at a given timestep we find that the occurrence of both events simultaneously and co-spatially is between 10 and 20\%. When these two phenomena do occur together, they do so nearly simultaneously (see e.g. the case described in Section~\ref{caso3}) or with a delay of a few minutes (as in Section~\ref{caso2}). Our observations (see Table~\ref{table1}) also show that whenever EBs and UV bursts coexist, the EB is always longer lived by several minutes, with the UV brightening appearing later (whether it is only a few seconds or a few minutes delay). The intensity evolution of the UV burst presents a more rapid rise and fall --a more impulsive behavior-- than its associated EB. In fact, the numerical simulations in Paper I point towards the same situation: the UV burst lights up, its temperature rises rapidly in the region between 500 and 3000 km and falls again after about two hundred seconds, while the EB exists for a longer period appearing before and disappearing after the UV burst. 
Both the light-curves and the time-sliced spectra, which reveal the spectral history of our magnetic reconnection cases, make evident a sequential nature of heating events. These figures show that EB and UV burst intensities evolve in unison, but there is a sequential behavior: first the \halpha~lits up, then the \MgII, and finally the \ion{Si}{IV}. We thus observe a clear temporal and spatial relationship between these kind of events whenever they coexist. In view of the results from Paper I, this could be interpreted as a reconnection sequence happening from the bottom to the top of the long current sheet or, alternatively, as an indication that it takes more time to heat the atmosphere to transition region temperatures than to heat the denser plasma closer to the photosphere to emit in \halpha. Another result revealed by the light-curves of the eight cases is that, in general, the intensity peaks of those EBs and UV bursts that are colocated and simultaneous are brighter than the peaks of those reconnection events (EB or UV burst) that occur independently of each other. In the same light-curves, all of the studied UV bursts show a bursty (meaning intermittent) behavior with sometimes up to five intensity peaks during the observation sequence.  

Our observations also show the existence of photospheric and chromospheric absorption lines superimposed on the broadened wings of our NUV and FUV IRIS lines. Specifically we have found \ion{Ni}{II} 139.3 nm superimposed at $-93$ \kms~on the wings of the \ion{Si}{IV} 139.4 nm profiles, as well as \ion{Mn}{I} superimposed in the blue wing of the \ion{Mg}{II} k profiles. Regarding \ion{Ni}{II}, it is visible in some of the UV bursts analyzed here and not in others, as the \ion{Si}{IV} profiles are not always broad enough to show this absorption. In those bursts with broad enough wings, as in the September 6 case, the absorption is detected during the whole lifetime of the UV brightening. Similarly, the \ion{Mn}{I} absorption is found in some of our UV brightenings but not in others (see Table~\ref{table1}). In those bursts whose \ion{Mg}{II} profiles present the \ion{Mn}{I} absorption, it is detected during some timesteps within the burst evolution but not in others (see, e.g., Fig.~\ref{fig_caso2_spectra+timespectra}). This variability seems to be related to the presence or absence of the surge's cold gas which, at times, partially covers the UV burst area. Similar absorption features on the \ion{Mg}{II} line have been reported by \cite{2015ApJ...811..137T}
due to the recurrent ejection of surges from light bridges. \citet{2016ApJ...824...96T} also observe such absorption lines superimposed on the IRIS lines in the data set they analyze. However, they attribute the presence of these absorption lines to the low formation height of their UV bursts. The argument at use is that one observes such absorption lines when cold gas is stacked above the UV brightening, and since the \ion{Mn}{I} line e.g. is formed in the upper photosphere, the cold gas cloud must be lying at that height and the UV burst must be formed below that layer. However, another reason for the presence of absorption lines superimposed on the broadened UV IRIS lines is possible. Both observations and simulations \citep{2009A&A...507..949T,2014ApJ...781..126O,2015ApJ...810..145D,2016ApJ...825...93O,2016ApJ...822...18N,2017ApJ...839...22H,2019A&A...626A..33H} describe a scenario where cool, dense, magnetized bubbles of gas rise slowly through the photosphere and the chromosphere, expanding both vertically and horizontally and filling the chromosphere and transition region. In the process the newly uplifted cool, dense gas pumps the necessary opacity into greater heights that explains absorption in lines such as \ion{Mn}{I} and \ion{Ni}{II}.

\citet{2016ApJ...824...96T} suggest that the \MgII~lines can be used similarly to \halpha~for the identification of EBs, because in their EB-related UV bursts the \MgII~wings have a significant brightening but no obvious brightening in the line core is observed. However, our EB-only case (described in Sect.~\ref{caso4}) shows no sign of brightening in the wings of \MgII, only a very weak increase in the core, and a marginal increase in the triplet only during the peak intensity of the EB, which would rule out the possibility of using \MgII~to trace EBs in the IRIS database. In addition, all our UV bursts leave an imprint in the \MgII~profiles, regardless of whether they have an associated EB or not. On the other hand, we notice strong similarities between \CaII~and \halpha~in EBs, because the wings of \halpha~and \CaII~follow a similar temporal evolution and even do so the line cores of these spectral lines. However, \citet{2013ApJ...774...32V} warn that most EBs are also observed in the wings of \CaII~but with a markedly different morphology than in \halpha.

Table~\ref{table1} provides evidence that all UV bursts present a surge associated with them. The only case in which we do not observe a surge is the EB-only case on September 5. Since all UV bursts cases occur at the site of merging of magnetic fields of opposite polarities, which in turn happen within a region of vigorous flux emergence, we can state that there is a causal relationship between flux emergence and the existence of surges and UV bursts. \citet{2017ApJ...850..153N} investigated the details of this relationship and how newly emerged fields interacting with the ambient field produce surges and UV bursts (with and without coronal response). The numerical simulation of Paper I also produces a surge when a UV burst coexists with an EB. Even though EBs are also caused by flux emergence and merging of opposite polarities, a clear EB-surge connection has not been established \citep[e.g.,][]{2013JPhCS.440a2007R,2013ApJ...774...32V} and our EB-only case, even though not statistically significant, points in the same direction. 

To complete the analysis reported in this work, we have studied the morphological and dynamical properties of one of our extended UV brightenings, examining in detail the variations of these properties across the burst area. While a few previous works \citep{2017ApJ...836...63T,2019A&A...627A.101V} present results on the positional dependence of the properties of bright points and reconnection events, these works do not carry out an extensive and detailed characterization of the variation of their properties with position. Here we take into account the whole extension of the UV brightening and consider the physical properties at each pixel within the burst domain. Different parts of the UV burst are characterized by different \ion{Si}{IV} profile shapes: single-peaked, double-peaked, or a blend of both, i.e., single-peaked with a lesser component in one of the wings. At the same time, when analyzing the velocities obtained from the \ion{Si}{IV} 140.3~nm spectra and their spatial distribution we realized that the burst domain has three different zones: a redshifted area closer to disk center of $1 \arcsec$ width, a blueshifted area in the center of the UV burst domain slightly less than $1 \arcsec$ wide, and an area mainly at rest on the right-hand side of the burst extension. The redshifted zone contains mostly double-peaked profiles while the central blueshifted zone contains mainly either single-peaked profiles or single-peaked profiles with a satellite component in one wing. Finally, the area where the \ion{Si}{IV} profiles are at rest are totally populated by single-peaked profiles. This fact could be due to projection effects where bidirectional flows (represented by double-peaked profiles) are better observed closer to disk center, or just due to the magnetic topology of each particular reconnection event.  Maximum \ion{Si}{IV} velocities reach blueshifts of $-50$~\kms\ and redshifts of 100~\kms. The highest velocities are attained before and during the peak of maximum intensity of the UV burst, decreasing afterwards. When comparing the morphology of the burst in various spectral features like the \MgII\ k2 and k3 peaks or \ion{Si}{IV} we also note significant differences in size and shape: the brightening in NUV is wider than in FUV wavelengths ($2 \arcsec$ width in the \MgII~triplet versus $1 \arcsec$ in \ion{Si}{IV}). We believe this is due to the decrease of plasma-$\beta$ with height: (overshooting) convective motions play a significant role near the photosphere, but the magnetic field is completely dominant at greater heights where the plasma densities are much lower and temperatures high enough to allow the emission of
transition region lines such as \ion{Si}{IV}.

The analysis of the morphological, spectral, temporal and dynamical properties of an extended UV burst yields the picture of a nearly vertical current sheet oriented in the $y$ direction of the FOV. Along this current sheet, non-stationary magnetic reconnection and plasmoids occur. As a result, turbulence and hot irregular bidirectional jets are produced, with downflows on the leftward side of the UV burst and upflows in the central-rightward area of the burst. These results agree with, complete and support the scenario set by the numerical simulations of Paper I. 

The works by \citet{2014Sci...346C.315P,2015ApJ...812...11V,2016A&A...593A..32G,2016ApJ...824...96T,2018ApJ...854..174T} situate the formation of UV brightenings very low in the solar atmosphere, even at photospheric levels, even though current EB models argue against it because the typical EB diagnostics cannot be formed if the photosphere is heated to such extreme transition region temperatures. \citet{2016ApJ...824...96T} conclude that some of their UV brightenings are not related to EBs, but other UV bursts would actually be indeed EBs because they also show the typical signatures of an EB. In Paper I and this work we propose an alternative scenario: reconnection due to flux emergence does not occur at discrete heights, but along a long (few thousand km) tall {\it wall} (in more technical words a current sheet) where heating, plasmoids, turbulence and bidirectional jets are produced. In this case coexisting EBs and UV bursts would be part of the same reconnection system, but happening far apart vertically. \citet{2017ApJ...839...22H} already proposed that EBs form in the photosphere and UV bursts in the mid-chromosphere around 1300 km higher up (except that this work did not consider co-spatial and co-temporal events but independent from each other). In such way the same pixel at, or close to disk center, can exhibit a typical EB behavior when observing in \halpha\ and a typical UV burst shape when observing in IRIS upper chromosphere and transition region spectra. The heating events happen at different heights and this is evidenced by each of the diagnostics.

The observations presented here do not allow us to conclude the exact formation height of UV bursts relative to EBs. That was one of the main results from the numerical experiments of Paper I. Here we evidence the clear temporal and spatial relationship between coexisting EBs and UV brightenings, who behave at unison with a certain sequential character. This work complements and gives observational evidence to Paper I, but also expands further the investigation of the properties of UV bursts. For the first time an extended UV burst has been analyzed in detail and important physical characteristics have been derived. These series of articles characterize the process of flux emergence, the interaction of the new field with itself and the ambient field, and scrutinizes the transient dynamical phenomena resulting from magnetic reconnection, namely the relationship between EBs and UV bursts.

\begin{acknowledgements}
The Swedish 1-m Solar Telescope is operated on the island of La Palma (Spain)
by the Institute for Solar Physics of Stockholm University in the
Spanish Observatorio del Roque de los Muchachos of the Instituto de
Astrof{\'\i}sica de Canarias. The Institute for Solar Physics is supported by a grant for research infrastructures of national importance from the Swedish Research Council (registration number 2017-00625).
This research is supported by the Research Council of Norway, project number 250810, and through its Centers of Excellence scheme, project number 262622.
We made much use of NASA's Astrophysics Data System Bibliographic Services.

\end{acknowledgements}

\bibliographystyle{aa} 
\bibliography{EB_UVbursts}

\end{document}